\documentclass[aps,prd,amsmath,floats,floatfix,superscriptaddress,nofootinbib,showpacs]{revtex4}

\usepackage{amssymb}
\usepackage{amsmath}
\usepackage{verbatim}
\usepackage{mathrsfs}
\usepackage{amsfonts}
\usepackage{latexsym}
\usepackage{epsfig}
\usepackage{epstopdf}
\usepackage{color}
\usepackage{graphicx}
\usepackage[caption=false]{subfig}
\usepackage{units}
\usepackage{soul}
\usepackage{tcolorbox}
\usepackage{multirow}

\usepackage{booktabs}
\usepackage{tikz}
\usetikzlibrary{tikzmark}

\usepackage[hidelinks]{hyperref}
\usepackage{xcolor}
\hypersetup{
    colorlinks,
    linkcolor={red!50!black},
    citecolor={blue!50!black},
    urlcolor={blue!80!black}
}

\newcommand{\Real}{\mathbb{R}}
\newcommand{\Complex}{\mathbb{C}}

\newcommand{\im}{\mbox{Im}}
\newcommand{\Tr}{\mbox{Tr}}

\usepackage{abraces}

\definecolor{msmlbs}{rgb}{0.8 , 1.0 , 1.0 }
 \definecolor{mslbs}{rgb}{0.75, 1.0 , 0.75}
  \definecolor{msbs}{rgb}{1.0 , 0.9 , 1.0 }
    \definecolor{bs}{rgb}{1.0 , 1.0 , 0.8 } 
   \definecolor{lbs}{rgb}{0.75, 0.9 , 1.0 }
  \definecolor{mlbs}{rgb}{1.0 , 0.9 , 0.75}

\begin{document}


\newcommand{\argelia}[1]{\textcolor{red}{{\bf Argelia: #1}}}
\newcommand{\dario}[1]{\textcolor{red}{{\bf Dario: #1}}}
\newcommand{\juanc}[1]{\textcolor{green}{{\bf JC: #1}}}
\newcommand{\juan}[1]{\textcolor{cyan}{{\bf Juan B: #1}}}
\newcommand{\alberto}[1]{\textcolor{blue}{{\bf Alberto: #1}}}
\newcommand{\miguela}[1]{\textcolor{red}{{\bf Miguel: #1}}}
\newcommand{\mm}[1]{\textcolor{orange}{{\bf MM: #1}}}
\newcommand{\OS}[1]{\textcolor{blue}{{\bf Olivier: #1}}}

\long\def\symbolfootnote[#1]#2{\begingroup%
\def\thefootnote{\fnsymbol{footnote}}\footnote[#1]{#2}\endgroup}


\title{Boson stars and their relatives in semiclassical gravity}

\author{Miguel Alcubierre}
\affiliation{Instituto de Ciencias Nucleares, Universidad Nacional Aut\'onoma de M\'exico,
Circuito Exterior C.U., A.P. 70-543, M\'exico D.F. 04510, M\'exico}

\author{Juan Barranco}
\affiliation{Departamento de F\'isica, Divisi\'on de Ciencias e Ingenier\'ias,
Campus Le\'on, Universidad de Guanajuato, Le\'on 37150, M\'exico}

\author{Argelia Bernal}
\affiliation{Departamento de F\'isica, Divisi\'on de Ciencias e Ingenier\'ias,
Campus Le\'on, Universidad de Guanajuato, Le\'on 37150, M\'exico}

\author{Juan Carlos Degollado}
\affiliation{Instituto de Ciencias F\'isicas, Universidad Nacional Aut\'onoma de M\'exico,
Apdo. Postal 48-3, 62251, Cuernavaca, Morelos, M\'exico}

\author{Alberto Diez-Tejedor}
\affiliation{Departamento de F\'isica, Divisi\'on de Ciencias e Ingenier\'ias,
Campus Le\'on, Universidad de Guanajuato, Le\'on 37150, M\'exico}

\author{Miguel Megevand}
\affiliation{Instituto de F\'isica Enrique Gaviola, CONICET. Ciudad Universitaria, 5000 C\'ordoba, Argentina}

\author{Dar\'io N\'u\~nez}
\affiliation{Instituto de Ciencias Nucleares, Universidad Nacional Aut\'onoma de M\'exico,
Circuito Exterior C.U., A.P. 70-543, M\'exico D.F. 04510, M\'exico}

\author{Olivier Sarbach}
\affiliation{Instituto de F\'isica y Matem\'aticas, Universidad Michoacana de San Nicol\'as de Hidalgo,
Edificio C-3, Ciudad Universitaria, 58040 Morelia, Michoac\'an, M\'exico}


\date{\today}

 
\begin{abstract}
We construct boson star configurations in quantum field theory using the semiclassical gravity approximation. Restricting our attention to the static case, we show that the
semiclassical Einstein-Klein-Gordon system for a {\it single real quantum} scalar field whose state describes the excitation of $N$ {\it identical particles}, each one corresponding 
to a given energy level, can be reduced to the Einstein-Klein-Gordon system for $N$ {\it complex classical} scalar fields. Particular consideration is given to the spherically 
symmetric static scenario, where energy levels are labeled by quantum numbers $n$, $\ell$ and $m$. When all particles are accommodated in the ground state $n=\ell=m=0$, one recovers 
the standard static boson star solutions, that can be excited if $n\neq 0$. On the other hand, for the case where all particles have fixed radial and total angular momentum
numbers $n$ and $\ell$, with $\ell\neq 0$, but are homogeneously distributed with respect to their magnetic number $m$, one obtains the $\ell$-boson stars, whereas when $\ell=m=0$ 
and $n$ takes multiple values, the multi-state boson star solutions are obtained. Further generalizations of these configurations are presented, including the multi-$\ell$ 
multi-state boson stars, that constitute the most general solutions to the $N$-particle, static, spherically symmetric, semiclassical real Einstein-Klein-Gordon system, in which the 
total number of particles is definite.  In spite of the fact that the same spacetime configurations also appear in multi-field classical theories, in semiclassical gravity they 
arise naturally as the quantum fluctuations associated with the state of a single field describing a many-body system. Our results could have potential impact on direct detection 
experiments in the context of ultralight scalar field/fuzzy dark matter candidates.
\end{abstract}




\maketitle



\section{Introduction}
\label{sec:introduction}

Boson stars are exotic objects made of bosons in which the gravitational force that pulls matter together is counterbalanced by the dispersive nature of a scalar field. 
They were first proposed in the late 1960s by Kaup~\cite{Kaup68} and Ruffini and Bonazzola~\cite{Ruffini:1969qy}, and since then they
have been actively studied for more than half a century~\cite{Colpi:1986ye,Friedberg87,Gleiser:1988rq,Lee:1988av,Bizon:2000,Guzman:2004wj,Chavanis:2011cz} 
---see~\cite{Jetzer:1991jr,Schunck:2003kk,Liebling:2012fv,Zhang:2018slz,Visinelli:2021uve} for reviews on boson stars
and~\cite{Lee:1991ax,Rajaraman,Manton2007,EJWeinberg,Shnir} for references on other soliton solutions. At present, boson stars remain largely theoretical, although they have been 
employed to describe dark compact objects~\cite{Torres:2000dw,Guzman:2005bs,Cardoso:2019rvt,Hogan:1988mp,Kolb:1993zz,Palenzuela:2017kcg} and galactic halo 
cores~\cite{Sin:1992bg,Schive:2014dra,Schive:2014hza,Schwabe:2016rze,Veltmaat:2016rxo,Du:2016aik,Gonzalez-Morales:2016yaf} in models of axion~\cite{Peccei77,Preskill83,Abbott83,Dine83,Hertzberg08,Sikivie08} 
and axion-like~\cite{Hu:2000ke,Matos:2000ss,Arvanitaki:2009fg,Suarez:2013iw,Marsh:2015xka,Hui:2016ltb,Niemeyer:2019aqm,Luis2019,Ferreira:2020fam} particles. 

More pragmatically, a boson star is a regular, localized
solution to the {\it classical} Einstein-Klein-Gordon (EKG) system.
Nevertheless, nature is {\it quantum} at a fundamental level, and as
such these objects must also allow an interpretation in quantum field
theory. The purpose of this paper is to construct boson star
configurations in semiclassical gravity, to catalog their spectrum of
spherically symmetric equilibrium solutions, and to compare them with
those of the classical theory. Previous attempts to construct semiclassical boson stars have been carried out
in~\cite{Ruffini:1969qy,Ho:2002vz,Matos2007,Bernal:2009zy,UrenaLopez:2010ur,Barranco:2010ib}. See also Refs.~\cite{Berczi:2020nqy,Berczi:2021hdh,Guenther:2020kro}
for an analysis of the semiclassical gravitational collapse of quantum
matter and~\cite{arrechea2021semiclassical} for a recent study on
fluid stars in semiclassical gravity.

The semiclassical theory of gravity is an effective description of
gravitational phenomena that deals with gravitons at tree level and with
matter fields at one
loop~\cite{Verdaguer2020,semiclassical,Jordan:1986ug,Calzetta:1986ey,Paz:1990jg,Campos:1993ug}.
The resulting equations of motion are those of quantum field theory on
curved spacetimes coupled to the semiclassical Einstein equations,
where the expectation value of the stress energy-momentum tensor
operator acts as the source term on its right-hand side. In this
article, we introduce a well-defined operational program that deals
with free quantum fields acting as sources of stationary
spacetimes. Our program relies on the {\it semiclassical
  self-consistent configurations} proposed in
Ref.~\cite{DiezTejedor2012}, and it can be summarized into three
steps:

{\it i}) Consider a stationary, globally hyperbolic background spacetime
on which the free quantum fields are defined. The assumption of
stationarity allows one to introduce a preferred space of
``positive-norm" solutions of the matter field equations, hence a
preferred vacuum state. This in turn provides a well-defined theory
for the quantum fields, which we describe in terms of a Fock space
representation. In particular, field operators can be written
(formally) as linear combinations of creation and annihilation
operators, wherein the ``coefficients'' are mode functions $f_I(x)$ that solve
the {\it classical complex} field equations.

{\it ii}) Compute the expectation value $\langle \hat{T}_{\mu\nu}\rangle$
of the stress energy-momentum tensor operator with respect to a given
state in the Fock space. In order to do so we need a
regularization and renormalization prescription that removes the
ill-defined ultraviolet behavior of the theory, leading to sensible
finite outcomes. To achieve this, in this work we impose normal
ordering. More sophisticated approaches include e.g. adiabatic
subtraction~\cite{Parker1974} and Pauli-Villars
renormalization~\cite{Pauli1940,Weinberg:2010wq,Armendariz-Picon:2019csc},
although we expect the differences between such methods and ours
to be suppressed in the limit of large occupation
numbers, as we consider in our configurations, which we also assume to
be far from the Planck scale.
More generally, we can also compute a statistical average by tracing
$\hat{T}_{\mu\nu}$ with a density operator. This offers the
interesting possibility of considering, for instance, thermal states
with a given temperature.

{\it iii}) Solve the semiclassical Einstein equations $G_{\mu\nu} = 8\pi G
\langle \hat{T}_{\mu\nu}\rangle$ sourced by the expectation value (or
statistical average) of the (renormalized) stress energy-momentum tensor. This step
takes into account the backreaction of the quantum fields on the
classical geometry.

Of course, one of the main difficulties of this approach is that the
exact spacetime geometry is not known {\it a priori} in step {\it i}, and it
has to be constructed in a self-consistent way together with the other
two steps. For the purpose of illustration we will restrict our
attention to the case in which matter consists of a single, free, minimally coupled real
scalar field, although more involved situations can also be explored,
including quantum fields of higher rank. (The case of a complex scalar
field is analyzed in an appendix.)  In order to simplify the analysis
we shall further concentrate on static, spherically symmetric
configurations, but a generalization of our formalism to describe
stationary and axisymmetric rotating objects should be possible.  In
particular, as we show, in the static case the semiclassical EKG
equations can be reduced to a system of self-gravitating classical
complex scalar fields with harmonic time-dependency of the form
$e^{-i\omega_I t}$, which leads to a non-linear multi-eigenvalue
problem for the frequencies $\omega_I$.
These ``classical'' fields arise from the mode functions $f_I(x)$ that appear 
in the decomposition of the field operator and represent the ``wave functions'' of individual particles in first quantization.
Such problems were treated in the Newtonian limit long ago; see,
for example the seminal work by Lieb~\cite{Lieb77} or
Ref.~\cite{KavianMuschler2015} for more recent work in this direction.

In the static case, the resulting semiclassical solutions can be
interpreted as describing equilibrium selfgravitating objects made of
bosons.  Specifically, we construct general boson star configurations
in spherical symmetry, for which the number of particles in the
different energy levels is definite.  These objects interpolate
between standard boson
stars~\cite{Kaup68,Ruffini:1969qy,Colpi:1986ye,Friedberg87,Gleiser:1988rq,Lee:1988av,Guzman:2004wj,Bizon:2000,Chavanis:2011cz,Jetzer:1991jr,Schunck:2003kk,Liebling:2012fv,Zhang:2018slz,Visinelli:2021uve},
whose particles all lie in the lowest possible energy configuration,
to more general situations where the particles are accommodated in
states with higher energy and angular momentum, which include
$\ell$-boson
stars~\cite{Alcubierre:2018ahf,Alcubierre:2019qnh,Alcubierre:2021mvs,Alcubierre:2021psa,Jaramillo:2022zwg}
and multi-state boson
stars~\cite{Matos2007,Bernal:2009zy,UrenaLopez:2010ur}, as well as new
configurations obtained in this article: multi-$\ell$
multi-state boson stars. As we show, these constitute the most general solutions to the static, spherically
symmetric, semiclassical real EKG system for which the total number of
particles is definite (we shall refer to these configurations as
$N$-particle systems in this paper), and encompasses the previous
boson star solutions reported in the literature. A family tree of
these solutions is provided in Table~\ref{table:family}, where we show
how they are connected to each other.  A relevant question is whether
or not such configurations in which particles populate not only the
ground state, but also higher energy levels, are stable. For instance,
it has been found that boson stars with $\ell=0$ including only
excited states are unstable~\cite{Lee:1988av}. However, as discussed
in~\cite{Matos2007,Bernal:2009zy,UrenaLopez:2010ur}, a possible
mechanism of stabilization is to have a suitable combination of
particles in the ground and higher energy states.\footnote{See
also~\cite{DiGiovanni:2021vlu} for a stabilization mechanism that
includes another type of matter.} These multi-state configurations
arise naturally within our semiclassical description.
See~\cite{jaramillo19,nambo21} for $\ell$-boson stars in the Newtonian
limit and also~\cite{Vicens:2018kdk,Lin:2018whl} for related
configurations which include particles in the excited states.

Before we proceed with the construction of these objects, some words
are needed regarding the relation between the classical and quantum
descriptions of boson stars. Classical fields emerge from quantum
theories in the limit when the quantum fluctuations become negligible.
This limit is manifest, for instance, in the case of coherent states,
that saturate the quantum uncertainty principle and lead to field
configurations in which quantum fluctuations are reduced to their
minimum. However, in the real scalar field theory coherent states are
not compatible with a static spacetime geometry, as we prove later.
This is not surprising, given that for a real scalar field there are
no static configurations in the classical theory~\cite{Seidel91}.
This can be traced back to the properties of the classical limit
itself. The existence of soliton solutions relies on the presence of
conserved charges~\cite{Coleman:1985ki,Lee:1991ax}
which allow localized field configurations whose energy per unit charge is less than in any other solutions, 
including those in which all the charge is radiated to infinity. 
In the classical field theory for a real scalar field there is no such
charge, which explains the absence of static solutions associated with
the coherent states.

The situation is different in the quantum theory, where the conserved charge is the particle number that remains constant if the configuration is static.\footnote{As
we show in appendix~\ref{app.energy-momentum}, if the spacetime is static, the particle number operator commutes with the Hamiltonian of the system, which is the generator of the time translations.}
Interestingly, despite the fact that the $N$-particle solutions describe many-body systems and appear as a result of quantum fluctuations 
(and, consequently, lie beyond the classicality of
boson stars based on path integral arguments discussed in, e.g., reference~\cite{Herdeiro:2022gzp}), they still have a counterpart in multi-field classical theories.
This is due to the relation that exists between the semiclassical EKG system describing a single real scalar
field in a quantum state with a definite number of particles $N$ and the EKG system for $N$ complex classical scalar fields. 
In this way, we show that boson stars and their relatives (i.e., the multi-$\ell$ multi-state boson stars) can be understood within our program by invoking a  single real quantum scalar field without the need of postulating the existence of a fixed number of independent complex classical scalar fields, like, 
for example, the number $2\ell+1$ in the construction of $\ell$-boson stars as originally required in~\cite{Alcubierre:2018ahf}.
Nevertheless, the interpretation of the solutions is different in the classical and the quantum limits, and the difference can be found in the role 
that the mode functions $f_I(x)$ play in the different regimes of the theory. On one side, in an $N$-particle state the resulting complex fields $f_I(x)$ are understood as the wave functions of the individual particles in first quantization, 
and they represent the many-body Hartree approximation~\cite{Hartree1928} of a system of $N$ particles that live in the mean gravitational field that they produce, where the $N$-particle
wave function is just the product of  one-particle wave functions~\cite{Guth:2014hsa}. On the other hand, if the state of the quantum field is coherent, the mode functions play the role of a classical field excitation $\sum_I[\alpha_If_I(x)+\alpha_I^*f_I^*(x)]$, and a description in terms of particles is not appropriate in this case, in the same way that a description in terms of photons is not suitable in classical electrodynamics. 
Table~\ref{quantum/classical} sketches the connection between the classical and the quantum regimes.
Bearing in mind the different regimes of the theory may be relevant for potential direct detection experiments, such as those carried out 
in~\cite{Graham2013,Carney:2019cio,Donohue:2021jbv,Tsai:2021lly}.
References~\cite{Perez:2005gh,Sudarsky:2009za,Aguirre:2015mva,Sikivie:2016enz,Hertzberg:2016tal,Allali:2020shm} delve on the 
discussion of the classical and the quantum regimes of a scalar field in different cosmological and astrophysical situations.

This paper is organized as follows. In
section~\ref{sec.theoretical.framework} we review the main ingredients
of quantum field theory on curved spaces and semiclassical gravity.
In section~\ref{Sec:Static} we focus on the static case and next, in
section~\ref{sec:static.sph.sym} we further specialize to the static
spherically symmetric situation. This leads to the main theoretical
result of this article, which is summarized in the semiclassical EKG
system of
Eqs.~(\ref{Eq:KGStaticSphSym}),~(\ref{Eq:EinsteinSphSym1}), and~(\ref{Eq:EinsteinSphSym2}). Remarkably, as a consequence of the semiclassical approach the
resulting system of equations includes as particular case the system
for $N$ classical complex fields. This constitutes the starting point
for the subsequent analysis of this paper.  Numerical solutions
presenting new configurations which arise naturally in our formalism,
including multi-$\ell$, multi-state, and multi-$\ell$ multi-state
boson stars are presented in section~\ref{Sec:Numerical}. Conclusions
are drawn in section~\ref{Sec:Discussion} and technical aspects of our
calculations are included in appendix~\ref{app.energy-momentum}.  In
appendix~\ref{sec.complex} we introduce the static, spherically
symmetric, semiclassical complex EKG system which, in addition to the
$N$-particle configurations, allows solutions sourced by coherent
states that are not static.

Our conventions are as follows. We use the mostly plus signature
convention for the spacetime metric, $(-,+,+,+)$, and to simplify the
notation, we work in terms of natural units for which
$\hbar=c=1$. Numerical results are obtained using Planck units, where
in addition we set $G=1$.

\begin{table}

\caption{Minimal ingredients to construct boson stars and their
  relatives in the different regimes of a quantum scalar field theory
  in the semiclassical gravity approximation.  In this paper we
  concentrate on static configurations.  When the quantum field is in
  a coherent state, the mode functions are related with the excitation
  of a classical field. In contrast, when the quantum field is in a
  $N$-particle state, the mode functions are related to the particle wave functions describing a many-body system.}
\label{quantum/classical}
\begin{center}
\begin{tabular}{l l l l l l l }
 \hline
 \hline
  Regime && State   && Boson stars && Multi-$\ell$ multi-state boson stars \\
 \hline
  classical field excitation   &&  coherent state    && one complex field && $N$ complex fields \\
  many-body system && $N$-particle state    && one real/complex field && one real/complex field \\
 \hline
 \hline
\end{tabular}
\end{center}

\end{table}

\section{Theoretical framework}
\label{sec.theoretical.framework}

The quantization of a free field on a curved, globally hyperbolic
spacetime is well understood. This program was initiated by Parker in
the late 1960s and developed further by Fulling, Ford and Wald,
among others (see e.g.~\cite{Birrell1984, Wald1994, Mukhanov2007,
  Parker2009} for relevant textbooks on this subject and references to the aforementioned original work). If in addition
the quantum fields act as a source of the spacetime metric, the
semiclassical theory of gravity~\cite{Verdaguer2020,semiclassical}
provides an effective description that combines the quantum nature of
matter with the classical behavior that gravity exhibits at
macroscopic scales.\footnote{The regime of applicability of
semiclassical gravity is an open question, due mainly to the fact that
we do not have access to a complete, satisfactory theory of quantum
gravity.  See the discussion in Ref.~\cite{Flanagan:1996gw},
section~IIA (and references therein), for a critical examination of
the origin of the semiclassical equations.}  In this section we review
the main ingredients of this construction.

\subsection{Quantum spin-0 fields in curved spaces}\label{sec.QFT}

For the following, we consider a globally hyperbolic spacetime
$(\mathcal{M},ds^2)$ which is foliated by three-dimensional Cauchy
hypersurfaces $\Sigma_t$.  In terms of the standard $3+1$
decomposition the spacetime metric is written as
\begin{equation}\label{metric}
 ds^2 = -(\alpha^2-\beta_j\beta^j)dt^2 + 2\beta_i dt dx^i +\gamma_{ij} dx^i dx^j .
\end{equation}
Here, $\alpha (x)$ is the lapse function, $\beta^j(x)$ is the shift vector,
and $\gamma_{ij}(x)$ is the induced metric on $\Sigma_t$, with $x =
(t,\vec{x})$ denoting a generic point in the spacetime manifold. Latin
indices $i,j,k,\ldots$ take natural values in the range from $1$ to
$3$ and are raised and lowered with the three-metric $\gamma_{ij}$,
e.g., $\beta_i =\gamma_{ij}\beta^j$.

At the classical level, a real free massive scalar field satisfies the
Klein-Gordon equation
\begin{equation}
 (\Box - {m_0}^2)\,\phi = 0,
\label{KG}
\end{equation}
where $\Box := g^{\mu\nu}\nabla_{\mu}\nabla_{\nu}$ is the curved
d'Alembertian operator in four dimensions, $g^{\mu\nu}$ is the inverse of
the spacetime metric, and $\nabla_{\mu}$ is the covariant derivative with
respect to this metric. The parameter $m_0$, which we assume to be
positive, denotes the inverse Compton length of the field
(that plays the role of the rest mass of the particles in the quantum
theory), and for simplicity a minimal coupling with gravity has been
considered.

For the quantization of the field $\phi$, one extends the space of
real classical solutions to the space of complex-valued classical
solutions of Eq.~(\ref{KG}). Let us call this space $X$ in the
following. Given two such solutions $\phi_1,\phi_2\in X$, one
introduces the four-current vector field (with $\nabla^\mu :=
g^{\mu\nu}\nabla_\nu$)
\begin{equation}
j^\mu(\phi_1,\phi_2) :=
-i\left[ \phi_1(\nabla^\mu\phi_2^*) - (\nabla^\mu\phi_1)\phi_2^* \right],
\label{current}
\end{equation}
which, by virtue of Eq.~(\ref{KG}), is divergence-free ($\nabla_\mu
j^\mu = 0$) and satisfies the symmetries $j^\mu(\phi_1,\phi_2) =
[j^\mu(\phi_2,\phi_1)]^* = -j^\mu(\phi_2^*,\phi_1^*)$.  Here and in
the following, $\phi_2^*(x)$ denotes the complex conjugate of
$\phi_2(x)$. The four-current~(\ref{current}) gives rise to an inner
product on $X$, defined as
\begin{equation}
 (\phi_1,\phi_2) := \int_{\Sigma_t} j^\mu n_\mu d\gamma
  = - i \int_{\Sigma_t}[\phi_1(\pounds_n\phi_2^*) - (\pounds_n\phi_1)\phi_2^*] d\gamma.
\label{scalar}
\end{equation}
In this equation $(n_{\mu}) = (-\alpha,0,0,0)$ is the future-directed
time-like unit normal covector field to the Cauchy hypersurfaces
$\Sigma_t$, $\pounds_n\phi_2 = n^\mu\nabla_\mu\phi_2$ refers to the
Lie derivative of $\phi_2$ with respect to the corresponding vector
field $n = g^{\mu\nu} n_\mu\partial_\nu$, and $d\gamma =
\sqrt{\det(\gamma_{ij})}\,d^3 x$ denotes the volume element on this
hypersurface.  As long as the space $X$ is restricted to those
solutions of Eq.~(\ref{KG}) which decay sufficiently fast at spatial
infinity, the inner product~(\ref{scalar}) does not depend on the
choice of the Cauchy hypersurface. Note also that by construction the
inner product~(\ref{scalar}) is linear in its first argument and
inherits the symmetries of the four-current, such that
$(\phi_1,\phi_2) = (\phi_2,\phi_1)^* = -(\phi_2^*,\phi_1^*)$. However,
it fails to be positive definite. Indeed, for $\phi_1 = \phi_2 =:
\phi\in X$,
\begin{equation}
(\phi,\phi) = 2\,\im \int_{\Sigma_t} \phi (\pounds_n\phi^*) d\gamma
\end{equation}
may assume any real (positive or negative) value, since the
restrictions of the functions $\phi$ and $\pounds_n\phi$ on $\Sigma_t$
represent the Cauchy data for Eq.~(\ref{KG}), which is free.

At the quantum level, the scalar field and its conjugate momentum
$\pi(x) := \sqrt{\det(\gamma_{ij})}\pounds_n\phi(x)$ are promoted to
self-adjoint field operators $\hat{\phi}(x)$ and $\hat{\pi}(x) =
\sqrt{\det(\gamma_{ij})}\pounds_n\hat{\phi}(x)$ acting on an abstract
Hilbert space $\mathscr{H}$. These operators satisfy the standard
equal time commutation relations
\begin{equation}\label{eq.commutators}
 [\hat{\phi}(t,\vec{x}),\hat{\pi}(t,\vec{y})] = i\delta^{(3)}(\vec{x}-\vec{y}),
\qquad
 [\hat{\phi}(t,\vec{x}),\hat{\phi}(t,\vec{y})] = [\hat{\pi}(t,\vec{x}),\hat{\pi}(t,\vec{y})] = 0,
\end{equation}
for all $(t,\vec{x}),(t,\vec{y})\in \Sigma_t$ and $\hat{\phi}(x)$
satisfying the Klein-Gordon equation~(\ref{KG}). Note that we follow a canonical quantization scheme, and we are
working in the Heisenberg representation, where the evolution is
codified in the operators and the state-vectors remain independent of
time. Instead of the field operator $\hat{\phi}(x)$, which is really
an operator-valued distribution on ${\cal M}$, it is sometimes
convenient to work with its ``smeared-out" versions, given by the
operators
\begin{equation}
\hat{a}(f) := (\hat{\phi},f),\quad f\in X,
\end{equation}
with $(\cdot,\cdot)$ the same inner product as in Eq.~(\ref{scalar}),
such that the definition is again independent of the choice of the
Cauchy surface and $\hat{a}(f)$ is constant in time.  Because
$\hat{\phi} = \hat{\phi}^\dagger$ is self-adjoint, it follows that
$\hat{a}^\dagger(f) = -(\hat{\phi},f^*) = -\hat{a}(f^*)$ and the
commutation relations~(\ref{eq.commutators}) imply
\begin{equation}
[\hat{a}(f),\hat{a}(g)^\dagger] = (g,f),\qquad
[\hat{a}(f),\hat{a}(g)] = -(g^*,f),
\label{commutator3}
\end{equation}
for all $f,g\in X$.

For the following, we make the important assumption that the space of
complex-valued, classical solutions $X$ of the Klein-Gordon
equation~(\ref{KG}) can be decomposed in the
form~\cite{Jacobson:2003vx}
\begin{equation}
X = X_+ \oplus X_+^*,
\label{decomposition}
\end{equation}
with the subspace $X_+$ consisting of solutions with positive norm
(that is, $(f,f) > 0$ for all $f\in X_+$ with $f\neq 0$) and its
complex conjugate $X_+^*$ being orthogonal to it, such that $(f,g^*) =
0$ for all $f,g\in X_+$.\footnote{Given the properties of the inner
product, the elements of $X_+^*$ have negative norm; however, this
will not be relevant for what follows.}  For a detailed discussion on
the validity and uniqueness of this decomposition we refer the reader
to~\cite{Ashtekar:1975zn}.  For a generic spacetime manifold $({\cal
M},ds^2)$, it does not seem clear if a split of this kind exists and is unique; 
  however, for static or stationary spacetimes, i.e., those
admitting a globally-defined timelike Killing vector field, the
decomposition of $X$ exists, and, moreover, the
``energy-requirement'' of Ref.~\cite{Ashtekar:1975zn} selects a
preferred one. Under these assumptions, the vacuum state (which in
general depends on the choice of the decomposition) can be
characterized as the state $\vert 0 \rangle \in \mathscr{H}$ for which
$\langle 0 \vert 0 \rangle = 1$ and $\hat{a}(f)\vert 0 \rangle = 0$
for all $f\in X_+$ in the preferred decomposition.  The particular
case for which $({\cal M},ds^2)$ is {\it static} will be reviewed in
the next section. In this case, the natural choice for $X_+$
satisfying the energy-requirement can be constructed directly from the
space of ``positive-frequency'' solutions of the Klein-Gordon
equation~(\ref{KG}).

To proceed, it is convenient to work with an orthonormal set of basis
functions $f_1,f_2,\ldots\in X_+$, such that $(f_I,f_J) = \delta_{IJ}$,
which are usually refereed to as
the {\it mode functions}, and
to introduce the corresponding creation and annihilation operators
\begin{equation}
\hat{a}_I := \hat{a}(f_I),\qquad
\hat{a}_I^\dagger := \hat{a}^\dagger(f_I),
\end{equation}
which, by virtue of Eq.~(\ref{commutator3}) and the
decomposition~(\ref{decomposition}) fulfill the commutation relations
\begin{equation}
[\hat{a}_I,\hat{a}^\dagger_J] = \delta_{IJ},\qquad
[\hat{a}_I,\hat{a}_J] = 0.
\label{commutator4}
\end{equation}
In terms of the operators $\hat{a}_I$ and $\hat{a}_I^\dagger$, the
field operator $\hat{\phi}(x)$ can be decomposed as
\begin{equation}
\hat{\phi}(x) = \sum_I\left[ \hat{a}_I f_I(x) +  \hat{a}^{\dagger}_I f^*_I(x) \right],
\label{expansion}
\end{equation}
which allows one to disentangle the field properties, codified in the
spacetime functions $f_I(x)$ and $f_I^*(x)$, from the time-independent
quantum operators $\hat{a}_I$ and $\hat{a}_I^\dagger$. Notice that
this decomposition is not unique, and any choice of the orthonormal
set of basis functions works equally well.

The Hilbert space $\mathscr{H}$ can now be constructed ({\it \`a la}
Fock) by successive applications of creation operators on the vacuum
state. A generic element in the base of the Fock construction can be
written in the form
\begin{equation}
 \vert N_1, N_2,\ldots\rangle = \frac{(\hat{a}^{\dagger}_1)^{N_1}}{\sqrt{N_1!}}\frac{(\hat{a}^{\dagger}_2)^{N_2}}{\sqrt{N_2!}}\ldots 
 \vert 0\rangle,
\label{FockBase}
\end{equation}
with $N_1, N_2, \ldots$ nonnegative integer numbers such that $\sum_I
N_I$ is finite. Using the commutation
relations~(\ref{commutator4}), one easily shows that the basis
vectors~(\ref{FockBase}) are normalized and mutually orthogonal and
that
\begin{subequations}
\begin{eqnarray}
\hat{a}_K^\dagger \vert N_1,\ldots N_K,\ldots\rangle  
 &=& \sqrt{N_K+1} \vert N_1,\ldots,N_K+1,\ldots\rangle,\\
 \hat{a}_K \vert N_1,\ldots, N_K, \ldots\rangle  
 &=& \sqrt{N_K} \vert N_1,\ldots,N_K-1,\ldots\rangle.
\end{eqnarray}
\end{subequations}
Furthermore, the states~(\ref{FockBase}) are eigenvectors of the
particle number operator $\hat{N}_K := \hat{a}^{\dagger}_K \hat{a}_K$,
such that $\hat{N}_K\vert N_1,\ldots, N_K,\ldots\rangle = N_K\vert
N_1,\ldots, N_K,\ldots\rangle$, with $N_K$ representing the number of
particles in the {\it K-th} mode, and thus the states~(\ref{FockBase})
describe a system of $N = \sum_I N_I$ identical quantum particles,
with $N_1$ of them in the one-particle state corresponding to mode
$1$, $N_2$ of them in the state corresponding to mode $2$, and so on.
Note that the expectation value of the field operator $\hat{\phi}(x)$
vanishes when evaluated on a state with a definite number of
particles, i.e., $\langle N_1,N_2,\ldots
|\hat{\phi}(x)|N_1,N_2,\ldots\rangle = 0$, although this does not
imply that the expectation value of the stress energy-momentum tensor
also vanishes, as we will see later.  Because the creation operators
$\hat{a}_I^\dagger$ commute with each other, these states are totally
symmetric and thus the particles satisfy the Bose-Einstein statistics
and describe bosons.

An arbitrary (pure) state in the Hilbert space $\mathscr{H}$ can be
expressed as a linear combination of the elements in the Fock
construction,
\begin{equation}
 \vert \psi\rangle = 
 \sum_{N_1,N_2,\ldots=0}^{\infty} 
 (C_{N_1N_2\ldots})\vert N_1,N_2,\ldots\rangle,
 \label{eq.state}
\end{equation}
with $C_{N_1N_2\ldots}$ arbitrary complex numbers such that
$\sum_{N_1,N_2,\ldots=0}^{\infty} \vert C_{N_1 N_2 \ldots}\vert^2=1$.
A case of particular interest consists of the coherent states, defined
as those elements of $\mathscr{H}$ that saturate the quantum
uncertainty principle and most closely resemble a classical field
excitation; see, e.g., page 97 of Ref.~\cite{Sakurai94} for a brief
description of the coherent states in the context of the
single-particle quantum harmonic oscillator
and~\cite{Zhang1990,Sanders2012} for comprehensive reviews. In the
context of a field theory, they are usually referred as Glauber
states~\cite{Glauber63} and are defined as the eigenstates of the
(non-Hermitian) annihilation operators $\hat{a}_I$,
\begin{equation}\label{eq.coherent}
 \hat{a}_K|\alpha_1,\alpha_2,\ldots\rangle = \alpha_K |\alpha_1,\alpha_2,\ldots\rangle,
\end{equation}
with $\alpha_K$ being in general complex numbers. Note that, contrary
to what happens for the states with a definite number of particles,
the expectation value of the scalar field does not vanish when
evaluated on a coherent state, where we obtain
$\langle\alpha_1,\alpha_2,\ldots|\hat{\phi}(x)|\alpha_1,\alpha_2,
\ldots\rangle = \sum_I [\alpha_I f_I(x)+\alpha_I^*f_I^*(x)]$, which is
a solution to the classical Klein-Gordon equation~(\ref{KG}). This is
not surprising and actually is a consequence of Ehrenfest's theorem
and the fact that we are dealing with a linear theory, so the
expectation value of the field operator $\hat{\phi}(x)$ always
satisfies the classical equations of motion.

\subsection{Semiclassical gravity}

So far, we have ignored the backreaction of the quantum fields, and we have assumed that the spacetime background is given {\it a priori}. However, according to general relativity, the spacetime
metric is determined dynamically by the distribution of matter through
Einstein's field equations, for which a (classical) stress
energy-momentum tensor is required.
One possibility to address this problem is to follow an effective
field theory approach where, starting from the generating functional
$Z[J,T^{\mu\nu}]$ (with $J$ and $T^{\mu\nu}$ external sources of
$\phi$ and $g_{\mu\nu}$), one expands the effective action
$\Gamma[\phi,g]$ at tree level in gravitons and one loop in matter
fields~\cite{Jordan:1986ug,Calzetta:1986ey,Paz:1990jg,Campos:1993ug}
(see also Ref.~\cite{Armendariz-Picon:2020tkc} for a derivation of the
quantum corrected equations of motion of the metric in terms of an
analysis of graviton fluctuations).  The resulting theory is known as
semiclassical gravity, where, in addition to the quantum field theory
summarized in section~\ref{sec.QFT},
one enforces Einstein's equations sourced by the expectation value of
the stress energy-momentum tensor~\cite{Verdaguer2020,semiclassical},
\begin{equation}
 G_{\mu\nu} = 8\pi G \langle\hat{T}_{\mu\nu}\rangle.
\label{Eins.semi}
\end{equation}
Here, $G_{\mu\nu}$ is the Einstein tensor, and
$\langle\hat{T}_{\mu\nu}\rangle =
\langle\psi\vert\hat{T}_{\mu\nu}\vert \psi\rangle$ denotes the expectation
value of the stress energy-momentum operator when evaluated on an
arbitrary state $\vert\psi\rangle$ in $\mathscr{H}$. Further details
on how to solve this problem based on the notion of semiclassical
self-consistent configurations introduced in
Ref.~\cite{DiezTejedor2012} will be given below. For recent
rigorous results on the initial-value problem for semiclassical gravity, see, for
instance, references~\cite{Juarez-Aubry:2019jon,bjuarez2021a,bjuarez2021b,Juarez-Aubry:2022qdp}.

For the case of a real free massive scalar field, the operator
associated with the stress energy-momentum tensor takes the form
\begin{equation}
\hat{T}_{\mu\nu} = 
(\nabla_{\mu}\hat{\phi})(\nabla_{\nu}\hat{\phi}) 
- \frac{1}{2}g_{\mu\nu}\left[ (\nabla_{\alpha}\hat{\phi})(\nabla^{\alpha}\hat{\phi})
 + {m_0}^2\hat{\phi}\hat{\phi} \right].
\label{energy.momentum}
\end{equation} 
It is important to notice that this quantity is quadratic in field
operators and that it contains products of these operators evaluated at the
same spacetime point. Since $\hat{\phi}$ is a distribution, these products are not
mathematically well-defined, and this problem manifests itself as
divergences when computing the right-hand side of
Eq.~(\ref{Eins.semi}). Some regularization and renormalization
prescription is needed in order to subtract the ill-defined
ultraviolet behavior from the expectation value of higher order
operators, providing sensible finite results.
On the one hand, this requires the introduction of counterterms into the
effective action, in such a way that the divergences that appear in
the free theory are absorbed into the cosmological constant, Newton's
gravitational constant, and the coupling constants accompanying
quadratic curvature scalars such as $R^2$ and
$R_{\mu\nu}R^{\mu\nu}$~\cite{semiclassical}
(which are expected to be suppressed in the low energy regime and we
do not include here).\footnote{The observed value of the cosmological
constant is so small that its relevance at local scales is negligible
and for that reason we will not include this term in our analysis
either.}
On the other hand, this also leads to a finite contribution to the
expectation value of the stress energy-momentum tensor originating
from the structure of the vacuum itself, that even if interesting in
its own right, will not be explored in more detail in the present
paper (this contribution is expected to be suppressed for large
occupation numbers, and this is what we assume in the following.) In
practice, this corresponds to assuming normal (Wick) ordering and
writing, e.g.,  $:\!\hat{a}_I \hat{a}_I^{\dagger}\!: \;=
\hat{a}_I^{\dagger} \hat{a}_I$ in our expressions,
moving all the creation operators to the left.

Introducing the field decomposition~(\ref{expansion}) in terms
of the creation and annihilation operators into the expression for the
stress energy-momentum tensor~(\ref{energy.momentum}), one obtains
\begin{equation}
\hat{T}_{\mu\nu} = 
\frac{1}{2}\sum_{I,J}\left[\hat{a}_I\hat{a}_J T_{\mu\nu}(f_I,f_J) 
 + \hat{a}^{\dagger}_I\hat{a}_J T_{\mu\nu}(f_I^*,f_J) + \textrm{H.c.} \right].
\label{energy.momentum.2}
\end{equation}
As usual, $\textrm{H.c.}$ stands for Hermitian conjugation, and to abbreviate
the notation we have defined
\begin{equation}
T_{\mu\nu}(f_I,f_J) := (\nabla_{\mu} f_I)(\nabla_{\nu} f_J) 
+ (\nabla_{\nu}f_I)(\nabla_{\mu} f_J) 
- g_{\mu\nu}\left[ (\nabla_{\alpha} f_I)(\nabla^{\alpha} f_J) + {m_0}^2 f_I f_J \right]  ,
\label{eq.T.f's}
\end{equation}
such that $T_{\mu\nu}(f_I,f_I^*)$ is the stress energy-momentum tensor
corresponding to a classical complex scalar field of amplitude
$f_I(x)$. With the normal order we have imposed,
Eq.~(\ref{energy.momentum.2}) provides sensible results. In
particular, for coherent states such as~(\ref{eq.coherent}) one has
$\langle\hat{a}_I\hat{a}_J\rangle=\alpha_I\alpha_J$ and
$\langle\hat{a}_I^\dagger\hat{a}_J\rangle=\alpha_I^*\alpha_J$, and the
expectation value of the stress energy-momentum tensor operator takes
the same form as its classical counterpart with
$\phi_{\textrm{cl}}(x)=\langle\hat{\phi}(x)\rangle = \sum_I [\alpha_I
  f_I(x)+\alpha_I^*f_I^*(x)]$, that is,
\begin{equation}
 \langle \alpha_1,\alpha_2,\ldots \vert \hat{T}_{\mu\nu} \vert \alpha_1,\alpha_2,\ldots\rangle 
   = \frac{1}{2} T_{\mu\nu}\left(\phi_{\textrm{cl}}, \phi_{\textrm{cl}} \right) ,
\label{eq.T.coherent}
\end{equation}
where the factor $1/2$ on the right hand side of
Eq.~(\ref{eq.T.coherent}) is due to the difference in the definition
of the stress energy-momentum tensor of a real and a complex field
[cf. Eqs.~(\ref{energy.momentum.2})
  and~(\ref{energy.momentum.complex})]. For a state with a definite
number of particles of the form~(\ref{FockBase}) we have,
however, $\langle\hat{a}_I\hat{a}_J\rangle=0$ and
$\langle\hat{a}_I^\dagger\hat{a}_J\rangle=N_I\delta_{IJ}$, and the
expectation value of the stress energy-momentum tensor reduces to
\begin{equation}
\langle N_1,N_2,\ldots \vert \hat{T}_{\mu\nu} \vert N_1,N_2,\ldots\rangle
= \sum\limits_I N_I T_{\mu\nu}(f_I,f_I^*)  ,
\label{expectation.value}
\end{equation}
which is also finite and equal to the weighted sum over the stress
energy-momentum tensors $T_{\mu\nu}(f_I,f_I^*)$ associated with each
mode function $f_I(x)$.  Note that there is no analog of
Eq.~(\ref{expectation.value}) in the classical real scalar field theory; this is because
the eigenstates~(\ref{FockBase}) of the particle number operator
satisfy $\langle\hat{\phi}(x)\rangle=0$, and in this case quantum
fluctuations
$[\langle\hat{\phi}^2(x)\rangle-\langle\hat{\phi}(x)\rangle^2]^{1/2}$
source the entire stress energy-momentum
tensor~(\ref{expectation.value}). Note also the different purpose that
the mode functions $f_I(x)$ serve in Eqs.~(\ref{eq.T.coherent})
and~(\ref{expectation.value}); whereas in the former expression they
are associated with the excitations of a classical field, in the latter,
they represent the wave functions of the quantum particles, which can be also interpreted as $N$ equal classical complex independent fields.

\subsection{Statistical ensembles}

Up to now, we have restricted our attention to pure states,
corresponding to rays in Hilbert space.  More generally, one
may consider a statistical ensemble described by a density operator
$\hat{\rho}$, that is, a self-adjoint nonnegative operator $\hat{\rho}
= \hat{\rho}^\dagger\geq 0$ of unit trace $\Tr(\hat{\rho}) = 1$. For
the particular case in which this operator is diagonal with respect to the
eigenvectors~(\ref{FockBase}) of the particle number operator,
$\hat{\rho}$ has the representation
\begin{equation}\label{eq.density.matrix}
\hat{\rho} = \sum\limits_{N_1,N_2,\ldots} (p_{N_1 N_2\ldots}) 
\vert N_1,N_2,\ldots \rangle\langle N_1, N_2, \ldots \vert  ,
\end{equation}
with the probabilities $0\leq p_{N_1 N_2\ldots}\leq 1$ satisfying
$\sum_{N_1,N_2,\ldots} (p_{N_1 N_2\ldots}) = 1$. Although this does
not describe the most general situation, it is sufficient to describe,
e.g., equilibrium systems at constant temperature $T$, in which case
the probabilities $p_{N_1 N_2\ldots}$ are subject to the Bose-Einstein
thermal equilibrium distribution. A more detailed analysis of such
thermal configurations lies beyond the scope of this article and will
be studied in future work.

When dealing with mixed states, one needs to replace the expectation
value that appears in the semiclassical Einstein
equations~(\ref{Eins.semi}) with the statistically averaged stress
energy-momentum tensor $\langle\hat{T}_{\mu\nu}\rangle_{\textrm{stat}}
= \Tr (\hat{\rho}\hat{T}_{\mu\nu})$, with $\Tr$ denoting the trace.
For a statistical ensemble of the
form~(\ref{eq.density.matrix}), it follows that
\begin{equation}
\Tr(\hat{\rho}\hat{T}_{\mu\nu}) 
= \sum\limits_I \langle N_I \rangle_{\textrm{stat}} T_{\mu\nu}(f_I,f_I^*) .
\label{stat.value}
\end{equation}
This has the same form as the right-hand side of Eq.~(\ref{expectation.value}), with $N_I$ replaced with its statistical
average
\begin{equation}\label{eq.N.statistical.average}
\langle N_I \rangle_{\textrm{stat}} := \sum\limits_{N_1,N_2,\ldots} (p_{N_1 N_2\ldots}) N_I  ,
\end{equation}
and the previous result~(\ref{expectation.value}) is recovered by
choosing all the $p_{N_1 N_2\ldots}$'s equal to zero except for
$p_{0,\ldots,N_I,0,\ldots} = 1$.

\subsection{Semiclassical self-consistent configurations}

In order to address a problem in semiclassical gravity it is
convenient to introduce the notion of {\it semiclassical
  self-consistent configurations}~\cite{DiezTejedor2012} (see also references~\cite{Juarez-Aubry:2022qdp,Canate:2018wtx}).
A semiclassical self-consistent configuration $\{\mathcal{M},
ds^2;\hat{\phi}(x),\hat{\pi}(x),\mathscr{H};|\psi\rangle\in\mathscr{H}\}$
consists of: {\it a}) a spacetime manifold $\mathcal{M}$ equipped with a metric $ds^2$, {\it b}) a quantum field theory $\hat{\phi}(x)$,
$\hat{\pi}(x)$ with the Hilbert space $\mathscr{H}$ defined on this fixed classical background geometry, and {\it c}) a state $|\psi\rangle$ in 
$\mathscr{H}$ such that the Klein-Gordon equation~(\ref{KG}) and
the semiclassical Einstein equations~(\ref{Eins.semi}) are satisfied
simultaneously at every point in the spacetime.  This is a
non-trivial task; in order to construct the Hilbert space
$\mathscr{H}$, we need to determine the subspace $X_+$ of positive norm
solutions $f_I(x)$ of the Klein-Gordon equation~(\ref{KG}), and these
solutions depend on the spacetime background which is obtained by
solving the semiclassical Einstein equations~(\ref{Eins.semi}), so
that both the metric field and the quantum state need to be determined
in a self-consistent way. In the case when the quantum theory consists
of a real scalar field, we will say that a semiclassical
self-consistent configuration constitutes a solution to the
semiclassical real EKG theory~(\ref{KG}) and~(\ref{Eins.semi}).

Having said this, we have identified three scenarios for which the
expectation value $\langle \hat{T}_{\mu\nu} \rangle$ of the stress
energy-momentum tensor operator has a special structure: {\it i}) coherent
states~(\ref{eq.coherent}), for which $\langle \hat{T}_{\mu\nu}
\rangle$ is equal to the stress energy-momentum tensor of the
corresponding classical solution
$\langle\hat{\phi}(x)\rangle= \sum_I [\alpha_I f_I(x) + \alpha_I^*
  f_I^*(x)]$, saturating the
quantum uncertainty principle;  {\it ii}) states with a definite number of
particles~(\ref{FockBase}), for which $\langle \hat{T}_{\mu\nu}
\rangle$ is sourced by quantum fluctuations and represents a weighted
sum over the classical stress energy-momentum tensors associated with
the complex fields $f_I(x)$; and  {\it iii}) statistical ensembles described by
a density operator $\hat{\rho}$ of the form~(\ref{eq.density.matrix}),
for which the statistical average $\Tr(\hat{\rho}\hat{T}_{\mu\nu})$
yields again a weighted sum of classical stress energy-momentum
tensors. In the first case, the semiclassical system is identical to
the classical EKG system for the {\it single, real}, free, minimally
coupled scalar field $\sum_I [\alpha_I f_I(x) + \alpha_I^*
  f_I^*(x)]$. In the second and third cases the semiclassical
equations are equivalent to the classical EKG system for a {\it family
  of non-interacting, complex}, free, minimally coupled scalar fields
$f_I(x)$ which need to form an orthonormal set of basis functions of
the subspace $X_+$ of positive norm solutions of the Klein-Gordon
equation. Note the different roles that the mode functions $f_I(x)$ play: in scenario  {\it i}, they combine into a single real field, whereas in scenarios  {\it ii} and  {\it iii}, 
they all constitute independent complex fields.
In this paper we concentrate mainly on scenarios  {\it ii} and  {\it iii}; however, we also
discuss scenario  {\it i} for the case of a complex scalar field in
appendix~\ref{sec.complex}.

Up to this point in the presentation, we have intended to provide the
reader with a general perspective of the problem of self-gravitating
boson systems in the semiclassical theory.  In the remainder of this
article, we focus on static configurations, in which case there is a
well-defined way of performing the
decomposition~(\ref{decomposition}).

\section{Static case}
\label{Sec:Static}

In addition to being globally hyperbolic, we now assume the spacetime
$({\cal M},ds^2)$ to be static, which implies that there exists a preferred
foliation ${\cal M} = \Real\times \Sigma$ of the spacetime manifold
such that the metric has the form
\begin{equation}
ds^2 = -\alpha^2(\vec{x}) dt^2 + \gamma_{ij}(\vec{x}) dx^i dx^j,
\label{Eq:StaticMetric}
\end{equation}
i.e., the shift vector is zero, $\beta^i = 0$, and the lapse function
$\alpha > 0$ and the induced three-metric $\gamma_{ij}$ only depend on
the spatial coordinates $\vec{x}$ on $\Sigma$.  Introducing this
ansatz into the Klein-Gordon equation~(\ref{KG}), we obtain
\begin{equation}\label{eq.KG2}
 \partial_t^2\phi -\alpha D^i\left( \alpha D_i \phi \right) + \alpha^2 {m_0}^2 \phi =0,
\end{equation}
where $\partial_t :=\partial/\partial t$ is the partial derivative
with respect to the time coordinate, and $D_i$ denotes the covariant
derivative operator associated with the induced metric $\gamma_{ij}$.
This equation contains no crossed terms of the form $\partial_t D_i$
and suggests the following ansatz for the basis functions,
\begin{equation}
f_I(t,\vec{x}) = \frac{1}{\sqrt{2\omega_I}}e^{-i\omega_I t}u_I(\vec{x}), 
\label{Eq:fIStatic}
\end{equation}
with $\omega_I > 0$ and $u_I$ a complex-valued function\footnote{Due
to the fact that the operator $H$ is real, we could in fact assume
that the functions $u_I$ are real-valued.  However, for later
convenience (see Sec.~\ref{sec:static.sph.sym}), we shall only assume
that the complex conjugate $u_I^*$ of $u_I$ is proportional to another
member of the same basis, which we call $u_{I'}$. Note that $\omega_I
= \omega_{I'}$.\label{Footnote:Five}} of the spatial coordinates
$\vec{x}$ only, and where the factor $1/\sqrt{2\omega_I}$ has been
introduced for future convenience.  In this case, the Klein-Gordon
equation~(\ref{eq.KG2}) leads to the following eigenvalue problem for
the square of the frequency ${\omega_I}^2$:
\begin{equation}
H u_I := -\alpha D^j\left( \alpha D_j u_I \right) + \alpha^2 {m_0}^2 u_I = {\omega_I}^2 u_I .
\label{Eq:KGStatic}
\end{equation}

The linear operator $H$ is formally self-adjoint on the Hilbert space
$Y$ of square-integrable functions $u: \Sigma\to \Complex$, with scalar
product
\begin{equation}
\langle u_1, u_2 \rangle := \int_\Sigma u_1^*(\vec{x}) u_2(\vec{x}) 
\frac{d\gamma} {\alpha(\vec{x})}  ,
\quad u_1,u_2\in Y .
\label{Eq:ScalarProd}
\end{equation}
Indeed, one can check that, for a suitable definition of the domain
$D(H)$ of the operator incorporating appropriate regularity and
fall-off conditions, we have $\langle u_1, H u_2\rangle =
\langle Hu_1 ,u_2 \rangle$ for all $u_1,u_2\in D(H)$. Furthermore,
\begin{equation}
\langle u, H u \rangle = \int_\Sigma \left( | Du(\vec{x})|^2 + {m_0}^2 |u(\vec{x})|^2 \right)
\alpha(\vec{x}) d\gamma,
\end{equation}
which is strictly positive for all $u\in D(H)$ different from
zero. Hence, $H$ is a symmetric positive operator, and since it
commutes with complex conjugation, Neumann's theorem (see Theorem X.3
in~\cite{ReedSimonVol2}) implies that it possesses a positive
self-adjoint extension. This offers the possibility of studying the
eigenvalue problem~(\ref{Eq:KGStatic}) using the powerful tools of
spectral theory for self-adjoint
operators~\cite{ReedSimonVol1,ReedSimonVol4,Much2021}.

In the following, we shall assume $H$ has a discrete spectrum with
corresponding eigenvalues $0 < \omega_1^2\leq \omega_2^2\leq \ldots$,
and associated eigenfunctions $u_1,u_2,\ldots$, which can be chosen
such that
\begin{equation}
\langle u_I,u_J \rangle = \delta_{IJ},\quad I,J = 1,2,\ldots   .
\label{Eq:uIOrtho}
\end{equation}
Each of these eigenfunctions gives rise to a (complex-valued) solution
of the Klein-Gordon equation of the form~(\ref{Eq:fIStatic}), which
together with its complex conjugate solution $f^*_I(t,\vec{x})$ can
easily be verified to satisfy the following properties:
\begin{equation}
(f_I,f_J) = -(f_I^*,f_J^*) = \delta_{IJ}  , \qquad
(f_I,f_J^*) = 0  ,
\end{equation}
where $(\cdot,\cdot)$ denotes the inner product defined in
Eq.~(\ref{scalar}).

If $H$ has a pure discrete spectrum, then the functions $u_I$ provide
an orthonormal basis for $Y$, and the functions $f_I$ and $f_I^*$
defined by Eq.~(\ref{Eq:fIStatic}) provide a basis of complex-valued
classical solutions of the Klein-Gordon equation, spanning the spaces
of ``positive-frequency" and ``negative-frequency" solutions,
respectively, which give rise to the spaces $X_+$ and $X_+^*$ of the
decomposition~(\ref{decomposition}).  Notice that this choice of the
decomposition makes essential use of the staticity of the spacetime,
i.e. the existence of the globally-defined hypersurface-orthogonal
timelike Killing vector field $\zeta = \partial_t$, where the
functions $f_I$ spanning the space $X_+$ have the property of being
eigenstates $\pounds_{\zeta} f_I = -i\omega_I f_I$ of 
$\zeta=\partial_t$ with eigenvalues $-i\omega_I$,
$\omega_I>0$ (see the ``energy-requirement'' of
Ref.~\cite{Ashtekar:1975zn} for details). If $H$ has a discrete and
continuous spectrum (as will be the case for the boson star solutions
discussed in the next section), the eigenfunctions $f_I$ are
incomplete; however, they may be completed by considering
``generalized" eigenfunctions lying outside the Hilbert space $Y$, as
it is usually done when dealing, for example, with free particles in
Minkowski space. Alternatively, one can also consider ``cutting off"
the spatial domain $\Sigma$ by replacing it with a compact subdomain
$\Sigma_R\subset \Sigma$ with a smooth outer boundary
$\partial\Sigma_R$ with large areal radius $R$, and by solving the
eigenvalue problem~(\ref{Eq:KGStatic}) on $\Sigma_R$ with homogeneous
Dirichlet conditions for $u_I$ on $\partial\Sigma_R$.  One then
obtains a pure discrete spectrum at the cost of introducing the
cut-off parameter $R$. However, as long as $R$ is much larger than the
size of the configuration, one would expect boundary effects to 
be negligible. Coming back to the case of free particles in Minkowski
space, this is what one usually does when introducing a fictitious box
of periodic boundary conditions and taking the limit of infinite
volume at the end of the calculation. For the configurations that we construct in this paper, only the discrete spectrum will be excited.

For a static configuration as described by
Eqs.~(\ref{Eq:StaticMetric}) and~(\ref{Eq:fIStatic}), the projections
of the semiclassical Einstein equations~(\ref{Eins.semi}) normal and
tangential to the hypersurface $\Sigma$ reduce to the system
[cf. Eqs.~(2.4.10) and~(2.5.4) in Ref.~\cite{Alcubierre2008}]
\begin{subequations}
\begin{eqnarray}
 R^{(3)} &=& 16\pi G\rho,\label{eq.einstein.static.Hamiltonian}\\
 R^{(3)}_{ij}-\frac{1}{\alpha}D_iD_j\alpha &=& 4\pi G \left[\gamma_{ij}(\rho-S)+2S_{ij} \right],\label{eq.einstein.static.dynamical}
\end{eqnarray}
\end{subequations}
where $R^{(3)}_{ij}$ and $R^{(3)}:=\gamma^{ij}R^{(3)}_{ij}$ refer to
the three-dimensional Ricci tensor and Ricci scalar with respect to
$\gamma_{ij}$, and
\begin{subequations}
\begin{eqnarray}
 \rho &:=& n^{\mu}n^{\nu}\langle\hat{T}_{\mu\nu}\rangle  , \label{eq.density.gen} \\
 S_{ij} &:=& (\delta_i{}^\mu + n_i n^\mu)(\delta_j{}^\nu + n_j n^\nu)\langle\hat{T}_{\mu\nu} \rangle  \label{eq.stress.gen}
\end{eqnarray}
are the expectation value of the energy density and the spatial stress
tensor as measured by the so-called Eulerian observers (those moving
along the normal direction to the spatial hypersurfaces), with $S :=
\gamma^{ij} S_{ij}$. For self-consistency with the staticity property,
$\rho$ and $S_{ij}$ also need to be time-independent, and the momentum
flux given by
\begin{equation}
 j_i :=  (\delta_i{}^\mu + n_i n^\mu) n^{\nu} \langle\hat{T}_{\mu\nu}\rangle,  \label{eq.flux.gen}
\end{equation}
\end{subequations}
must vanish.  For an arbitrary state in $\mathscr{H}$, the energy
density, the momentum flux, and the spatial stress tensor can be
expressed in the form~(\ref{eqs.general}) of
appendix~\ref{app.energy-momentum}.  As shown in this appendix,
$\rho$, $j_i$ and $S_{ij}$ are time-independent as long as
$\langle\hat{a}_I\hat{a}_J\rangle = 0$ for all $I,J$, and
$\langle\hat{a}^{\dagger}_I\hat{a}_J\rangle = 0$ whenever
$\omega_I\neq\omega_J$.
These conditions cannot be fulfilled for a non-trivial coherent state
like in Eq.~(\ref{eq.coherent}), where
$\langle\hat{a}_I\hat{a}_J\rangle = \alpha_I\alpha_J $ is different
from zero at least for some values of $I$ and $J$. However, for a
state with a definite number of particles as in Eq.~(\ref{FockBase}),
it follows that $\langle\hat{a}_I\hat{a}_J\rangle =0$ and
$\langle\hat{a}^{\dagger}_I\hat{a}_J\rangle = N_I \delta_{IJ}$, and we
obtain
\begin{subequations}\label{eq.T.coherent.state}
\begin{eqnarray}
\rho &=& \sum_I  \frac{N_I}{2\omega_I}\left[ | Du_I|^2+
\left( \frac{{\omega_I}^2}{\alpha^2} + {m_0}^2 \right) |u_I|^2 \right] ,\\
j_k &=& \sum_I \frac{N_I}{2}\frac{i}{\alpha}\left[ (D_k u_I) u_I^* - u_I(D_k u_I^*) \right] ,\\
S_{ij} &=& \sum_I \frac{N_I}{2\omega_I} \left\{ (D_iu_I)(D_ju_I^*) + (D_ju_I)(D_iu_I^*)
- \gamma_{ij}\left[|D_iu_I|^2-\left(\frac{{\omega_I}^2}{\alpha^2}-{m_0}^2\right)|u_I|^2\right]
\right\} ,
\end{eqnarray}
\end{subequations}
where we have abbreviated $|D u_I|^2 := \gamma^{ij} (D_i u_J)(D_j
u_J^*)$, and where we recall that the functions $u_I$ are subject to
the orthogonality condition $\langle u_I,u_J\rangle = \delta_{IJ}$. If
all the $u_I$'s are chosen to be real-valued, the momentum flux
obviously vanishes. More generally, if $u_I$ is complex-valued,
following the convention in footnote~\ref{Footnote:Five}, $j_k = 0$
follows provided that $N_I = N_{I'}$.  See
appendix~\ref{app.energy-momentum} for further information regarding
these conditions and their necessity in the context of static and
stationary states.

Taking into account Eqs.~(\ref{eq.T.coherent.state}), and imposing
$j_k=0$, the Hamiltonian
constraint~(\ref{eq.einstein.static.Hamiltonian}) and the trace and
traceless parts of Eq.~(\ref{eq.einstein.static.dynamical}) yield
\begin{subequations}\label{Eq:EinsteinStatic}
\begin{eqnarray}
 R^{(3)} &=& 8\pi G\sum\limits_I \frac{N_I}{\omega_I} \left[ | Du_I|^2 +
\left( \frac{{\omega_I}^2}{\alpha^2} + {m_0}^2 \right) |u_I|^2  \right] ,
\label{Eq:EinsteinStatic2} \\
 \frac{D^j D_j\alpha}{\alpha} &=& 8\pi G\sum\limits_I \frac{N_I}{\omega_I}
\left[\left( \frac{{\omega_I}^2}{\alpha^2} - \frac{{m_0}^2}{2} \right) |u_I|^2\right] ,
\label{Eq:EinsteinStatic1} \\
 \left[ R^{(3)}_{ij} - \frac{1}{\alpha} D_i D_j\alpha \right]^{\rm tf} 
 &=& 4\pi G\sum\limits_I \frac{ N_I }{\omega_I}
\left[ (D_i u_I)(D_j u_I^*) + (D_j u_I) (D_i u_I^*) \right]^{\rm tf} ,
\label{Eq:EinsteinStatic3}
\end{eqnarray}
\end{subequations}
where the superscript ``tf'' refers to the trace-free part with
respect to $\gamma_{ij}$, i.e. $(A_{ij})^{\rm tf} :=
A_{ij}-\frac{1}{3}\gamma_{ij}(\gamma^{mn}A_{mn})$.  These equations,
together with the Klein-Gordon equation~(\ref{Eq:KGStatic}),
constitute a nonlinear multi-eigenvalue problem for the frequencies
$\omega_I$ describing a system of $N= \sum_I N_I$ identical quantum
particles in self-gravitating equilibrium. Note that
Eqs.~(\ref{Eq:EinsteinStatic}) are also applicable to systems that are
described in terms of a statistical ensemble of the
form~(\ref{eq.density.matrix}); in this case, $N_I$ needs to be
replaced with its statistical average $\langle
N_I\rangle_{\textrm{stat}}$, see Eq.~(\ref{eq.N.statistical.average}).

In the next section, we further specialize these equations to the
static, spherically symmetric case, and we show that for states of
definite number of particles, as well as for statistical ensembles of
the form~(\ref{eq.density.matrix}), the resulting equations give rise
to the $\ell$-boson star configurations constructed
in~\cite{Alcubierre:2018ahf} (see
also~\cite{Alcubierre:2019qnh,Alcubierre:2021mvs}), and even to more
general solutions, a few examples of which are constructed numerically
in section~\ref{Sec:Numerical}.

\section{Static spherically symmetric configurations}
\label{sec:static.sph.sym}

We now further specialize to a static, spherically symmetric
spacetime, for which the three-metric can be expressed in the form:
\begin{equation}
\gamma_{ij} dx^i dx^j = \gamma^2 dr^2 + r^2 d\Omega^2,\qquad
\gamma = \left( 1 - \frac{2GM}{r} \right)^{-1/2},
\label{Eq:SphSym3Metric}
\end{equation}
where $M$ denotes the Misner-Sharp mass function and $d\Omega^2 =
d\vartheta^2+\sin^2\vartheta d\varphi^2$ is the standard line-element on
the unit two-sphere $S^2$.  Furthermore, in these coordinates, the
lapse $\alpha$, the function $\gamma$ and the Misner-Sharp mass $M$
only depend on the areal radius coordinate $r$. Because of the spherical
symmetry, the mode solutions of the Klein-Gordon
equation~(\ref{Eq:KGStatic}) can be assumed to be of the form
\begin{equation}
u_I(\vec{x}) = v_{n\ell}(r) Y^{\ell m}(\vartheta,\varphi) ,
\quad I = (n\ell m) ,
\label{Eq:SphSymAnsatz}
\end{equation}
with $Y^{\ell m}$ denoting the standard spherical harmonics, and where
no sum in the total angular momentum number $\ell$ is considered. Note
that $u_I(\vec{x})^* = (-1)^m u_{I'}(\vec{x})$ with $I' =
(n,\ell,-m)$, such that the property assumed in
footnote~\ref{Footnote:Five} is satisfied. Since the magnetic number
$m$ does not appear explicitly in the radial differential equation
\begin{equation}
 -\frac{\alpha}{\gamma r^2}\left( \frac{\alpha r^2}{\gamma} v'_{n\ell}\right)'
 + \alpha^2\left[\frac{\ell (\ell+1)}{r^2}+{m_0}^2\right] v_{n\ell} = {(\omega_{n\ell})}^2 v_{n\ell}
\label{Eq:KGStaticSphSym}
\end{equation}
that is obtained from Eq.~(\ref{Eq:KGStatic}) with the
ansatz~(\ref{Eq:SphSymAnsatz}), the radial functions $v_{n\ell}(r)$
can be chosen to be independent of this number.  Therefore, the
eigenvalue problem~(\ref{Eq:KGStatic}) reduces to finding (for each
$\ell = 0,1,2,\ldots$) a set of suitable radial basis functions
$v_{n\ell}$ solving Eq.~(\ref{Eq:KGStaticSphSym}). Using the
orthonormality property of the spherical harmonics, \mbox{$\int_{S^2}
  Y^{\ell m} Y^{\ell ' m'*} d\Omega =
  \delta_{\ell\ell'}\delta_{mm'}$}, the orthogonality condition
$\langle u_I, u_J \rangle = \delta_{IJ}$ reduces to
\begin{equation}
\int_0^{\infty} v_{n\ell}(r) v^*_{n' \ell}(r) \frac{\gamma(r)}{\alpha(r)} \: r^2 dr
= \delta_{nn'} .
\label{Eq:normalization}
\end{equation}
Assuming the functions $\alpha$ and $\gamma$ are regular at the center
$r = 0$, such that they have local expansions of the form $\alpha(r) =
\alpha_0 + \alpha_2 r^2 + \ldots$ and $\gamma(r) = 1 + \gamma_2 r^2 +
\ldots$, one can show that the local solution that is finite at $r =
0$ has the form $v_{n\ell}(r)\sim r^{\ell}$; see
Ref.~\cite{Alcubierre:2018ahf}.  Likewise, as $r\to \infty$, we impose
that the metric functions $\alpha$ and $\gamma$ converge to 1, and
that $v_{n\ell}(r)$ are bounded, which implies that they have the form
$v_{n\ell}(r) \sim e^{-\sqrt{{m_0}^2 - {(\omega_{n\ell})}^2} r}$, with
$0 < \omega_{n\ell} < m_0$.  A further restriction arises from the
identity $\langle u_I,H u_I \rangle = {\omega_I}^2$, which yields
\begin{equation}
\int_\Sigma \left\{ |D u_I|^2 + \left[ {m_0}^2
- \frac{{\omega_I}^2}{\alpha^2} \right] |u_I|^2 \right\} \alpha d\gamma = 0 ,
\label{Eq:Identity}
\end{equation}
and shows that ${m_0}^2 - {\omega_I}^2/\alpha^2$ cannot be positive
everywhere, since otherwise it would follow from
Eq.~(\ref{Eq:Identity}) that $u_I = 0$.

The functions $\alpha$ and $\gamma$ must be determined by solving the
static semiclassical Einstein field
equations~(\ref{Eq:EinsteinStatic}), where for consistency the
right-hand side must be a spherically symmetric tensor.\footnote{A
different approach to achieving this property was considered
in~\cite{Mielke:1980sa}, where the stress energy-momentum tensor is
averaged over the spheres in order to get rid of the angular
dependency. Our approach requires no such averaging.} This is clearly
the case if only the ground state is populated, i.e., if $N_I = 0$ for
all $I\neq (000)$, which gives rise to the standard boson star
equations. More generally, one can demand that $N_{n\ell m} = 0$ for
all $\ell > 0$, meaning that all the particles have zero angular
momentum but may nevertheless be in excited energy states. This gives
rise to the multi-state boson star equations solved
in~\cite{Matos2007,Bernal:2009zy,UrenaLopez:2010ur}.  Following the
same arguments as in appendix~A of Ref.~\cite{Alcubierre:2018ahf}, in
order for the expectation value of the stress energy-momentum tensor
to be spherically symmetric, it is in fact sufficient to choose
$N_{n\ell m} $ independent of the magnetic number $m$, such that
\begin{equation}
 N_{n,\ell, -\ell}  = N_{n,\ell, -(\ell-1)} =\ldots= N_{n,\ell, (\ell-1)} = N_{n,\ell,\ell},
\label{Eq:SphSymDistribution}
\end{equation}
which implies that the total angular momentum vanishes, even if the
individual particles posses angular momentum.  Note that this choice
also guarantees that the momentum flux is zero, as required for
staticity.  In other words, the excitation numbers $N_{n\ell m}$ are
functions of the energy levels $n$ and the total angular momentum
$\ell$, but not of the magnetic quantum number $m$. This is rather
similar to the case of a kinetic gas, in which a one-particle
distribution function depending only on the energy and the total
angular momentum gives rise to a static, spherically symmetric
configuration (see, for instance, section 5.1
in~\cite{Andreasson-LivRev}).

Assuming the validity of condition~(\ref{Eq:SphSymDistribution}) and
the spherically symmetric ansatz~(\ref{Eq:SphSym3Metric}),
equations~(\ref{Eq:EinsteinStatic}) reduce to
\begin{subequations}\label{Eq:EinsteinSphSym}
\begin{eqnarray}
 \frac{2GM'}{r^2} &=& \sum\limits_{n\ell}\frac{\kappa_{\ell}N_{n\ell m}}{\omega_{n\ell}}
 \left[ \frac{|v_{n\ell}'|^2}{\gamma^2} 
+ \left( \frac{{(\omega_{n\ell})}^2}{\alpha^2} + {m_0}^2 + \frac{\ell(\ell+1)}{r^2} \right) |v_{n\ell}|^2
\right] ,
\label{Eq:EinsteinSphSym1} \\
\frac{1}{\alpha\gamma r^2}\left( \frac{r^2\alpha'}{\gamma} \right)'
 &=& \sum\limits_{n\ell} \frac{\kappa_{\ell}N_{n\ell m}}{\omega_{n\ell}} \left[ \left(2\frac{{(\omega_{n\ell})}^2}{\alpha^2}
- {m_0}^2\right)|v_{n\ell}|^2\right] ,
 \label{Eq:EinsteinSphSym3} \\
\frac{(\alpha\gamma)'}{r\alpha\gamma^3} &=& \sum\limits_{n\ell}\frac{\kappa_{\ell}N_{n\ell m}}{\omega_{n\ell}} \left[ 
\frac{|v_{n\ell}'|^2}{\gamma^2} + \frac{{(\omega_{n\ell})}^2}{\alpha^2} |v_{n\ell}|^2 \right]  ,
\label{Eq:EinsteinSphSym2}
\end{eqnarray}
\end{subequations}
with $\kappa_{\ell}:= (2\ell+1)G$, and where the last identity was
obtained by contracting the angular components of
Eq.~(\ref{Eq:EinsteinStatic3}) with the metric of the unit two-sphere,
$\hat{g}_{AB}$, and making use of Eq.~(\ref{Eq:EinsteinSphSym1}). In
addition, we have also used the identities
$\sum_{m=-\ell}^{\ell}Y^{\ell m}Y^{\ell m*}=\frac{1}{4\pi}(2\ell+1)$
and $\sum_{m=-\ell}^{\ell}(\hat{\nabla}_A Y^{\ell m})(\hat{\nabla}^A
Y^{\ell m*})=\frac{1}{4\pi}\ell(\ell+1)(2\ell+1)$, where
$\hat{\nabla}_A$ makes reference to the covariant derivative with
respect to $\hat{g}_{AB}$ (see appendix~A in~\cite{Alcubierre:2018ahf}
for details). The full system of reduced static, spherically
symmetric, semiclassical real EKG equations consists of
Eqs.~(\ref{Eq:KGStaticSphSym}) and~(\ref{Eq:EinsteinSphSym}), where,
due to the twice contracted Bianchi identities,
Eq.~(\ref{Eq:EinsteinSphSym3}) can be omitted. Further, the
eigenfunctions $v_{n\ell}$ should satisfy the normalization
condition~(\ref{Eq:normalization}), although we can also do without
this equation if we absorb the occupation numbers $N_{n\ell m}$ in the
radial functions $v_{n\ell}$, as described in the next section. Once
the functions $v_{n\ell}$ are known, the quantum field $\hat{\phi}(x)$
can be reconstructed using Eqs.~(\ref{expansion}),~(\ref{Eq:fIStatic}), and~(\ref{Eq:SphSymAnsatz}), which gives
\begin{equation}
\hat{\phi}(x) = \sum\limits_{n \ell m} \frac{1}{\sqrt{2\omega_{n\ell}}}
\left[ \hat{a}_I e^{-i\omega_{n\ell} t} v_{n\ell}(r) Y^{\ell m}(\vartheta,\varphi) + \textrm{H.c.} \right].
\end{equation}

The $\ell$-boson star configurations we have discussed in
Refs.~\cite{Alcubierre:2018ahf,Alcubierre:2019qnh,Alcubierre:2021mvs}
are obtained by solving a particular case of this system, in which all
the $N_{n\ell m}$'s vanish except the ones for $n = 0$ and some
specific value of $\ell$. In this case, after absorbing the factor
$N_{0\ell m}/\omega_{0\ell}$ into the amplitude of $v_{0\ell}$, the
system of
Eqs.~(\ref{Eq:EinsteinSphSym1}),~(\ref{Eq:EinsteinSphSym2}), and~(\ref{Eq:KGStaticSphSym})
reduces precisely to the system~(7a),~(7b), and~(7c)
of~\cite{Alcubierre:2018ahf}.  However, in contrast to the purely
classical description
in~\cite{Alcubierre:2018ahf,Alcubierre:2019qnh,Alcubierre:2021mvs},
which requires precisely $2\ell+1$ complex scalar fields, the
semiclassical interpretation of the $\ell$-boson stars becomes much
more natural: they correspond to a particular excitation of a single 
real quantum spin zero field that describes a
selfgravitating system of $(2\ell+1)N_{0\ell m}$ identical quantum
particles of definite energy $E=\omega_{0\ell}$ and angular momentum
$L=\sqrt{\ell(\ell+1)}$ (both evaluated in natural
units). Furthermore, as is evident from the equations above, there are
many other possible configurations that can be constructed in this
way, involving excitations of different energy levels $n$ and
different total angular momentum numbers $\ell$. We summarize these
more general solutions and their sub-families, as well as a few
references to corresponding Newtonian configurations, in
Table~\ref{table:family}. Numerical examples of some of these more
general configurations are constructed in the next section.

\begin{table}[]\centering
\caption{
  Classification of the solutions to the static, spherically                                                                                                             
  symmetric, semiclassical real EKG system.  They represent                                                                                                            
  self-gravitating equilibrium configurations of a definite number of                                                                                                  
  identical quantum particles.  These are all the cases obtained when                                                                                                  
  combining two options for the radial quantum number $n$                                                                                                              
  and three options for the total angular momentum number $\ell$.                                                                                                      
  The two options for $n$ are: {\it i})~multiple values and {\it ii})~one value;                                                                                       
  while the three options for $\ell$ are: {\it i})~multiple values,                                                                                                    
  {\it ii})~one value, and {\it iii})~fixed value $\ell=0$. Note that                                                                                                  
  some of these solutions are particular cases of                                                                                                                      
  others. In order to show this hierarchy more clearly, we enclose                                                                                                     
  with a bracket solutions that are included in a more general                                                                                                         
  solution, which is indicated with an arrow.}

\begin{tabular}{r r r r r r l l l l l }
  \hline \hline                                                                                                                                                                                                                                         
                                &                &                &                &                & ~              & Name                                & $n$                      & $\ell$                          & Relativistic                                                                                  & Newtonian                         \\
                                                                                      \hline 
\tikzmark{p11}                  &                &                &                &                & \tikzmark{p10} & Multi-$\ell$ multi-state boson star & $n_1, n_2, \dotsc , n_p$ & $\ell_1, \ell_2,\dotsc, \ell_q$ & Sec.~\ref{Sec:Numerical}                                                                      & $\ldots$                          \\
                                & \tikzmark{p21} &                &                &                & \tikzmark{p20} & Multi-state $\ell$-boson star       & $n_1, n_2, \dotsc , n_p$ & $\ell_1$                        & Sec.~\ref{Sec:Numerical}                                                                      & $\ldots$                          \\ 
                                &                & \tikzmark{p31} &                &                & \tikzmark{p30} & Multi-state boson star              & $n_1, n_2, \dotsc , n_p$ & $0$                             & \cite{Matos2007,Bernal:2009zy,UrenaLopez:2010ur}                                              & \cite{KavianMuschler2015}         \\
                                &                & \tikzmark{p32} & \tikzmark{p62} & \tikzmark{p52} &                & Boson star                          & $n_1$                    & $0$                             & \cite{Kaup68,Ruffini:1969qy,Colpi:1986ye,Friedberg87,Gleiser:1988rq,Lee:1988av,Guzman:2004wj} & \cite{Lieb77,MorozPenroseTod1998} \\
                                & \tikzmark{p22} &                &                & \tikzmark{p51} & \tikzmark{p50} & $\ell$-Boson star                   & $n_1$                    & $\ell_1$                        & \cite{Alcubierre:2018ahf,Alcubierre:2019qnh,Alcubierre:2021mvs,Alcubierre:2021psa}            & \cite{jaramillo19,nambo21,GU2020}        \\
\tikzmark{p12}                  &                &                & \tikzmark{p61} &                & \tikzmark{p60} & Multi-$\ell$ boson star             & $n_1$                    & $\ell_1, \ell_2,\dotsc, \ell_q$ & Sec.~\ref{Sec:Numerical}                                                                      & \cite{GU2020}                     \\ 
  \hline \hline 
\end{tabular}
 
\begin{tikzpicture}[overlay,remember picture]
  \draw[-,black] ([yshift=0.8em]pic cs:p11) -- (pic cs:p12);
  \draw[-,black] ([yshift=0.8em]pic cs:p21) -- (pic cs:p22);
  \draw[-,black] ([yshift=0.8em]pic cs:p31) -- (pic cs:p32);
  \draw[-,black] (pic cs:p51) -- ([yshift=0.8em]pic cs:p52);
  \draw[-,black] (pic cs:p61) -- ([yshift=0.8em]pic cs:p62);
  \draw[<-,black] ([yshift=0.6ex]pic cs:p10) -- ([yshift=0.6ex]pic cs:p11);
  \draw[<-,black] ([yshift=0.6ex]pic cs:p20) -- ([yshift=0.6ex]pic cs:p21);
  \draw[<-,black] ([yshift=0.6ex]pic cs:p30) -- ([yshift=0.6ex]pic cs:p31);
  \draw[<-,black] ([yshift=0.6ex]pic cs:p50) -- ([yshift=0.6ex]pic cs:p51);
  \draw[<-,black] ([yshift=0.6ex]pic cs:p60) -- ([yshift=0.6ex]pic cs:p61);
  \draw[-,black] ([xshift=0.2em,yshift=0.8em]pic cs:p11) -- ([yshift=0.8em]pic cs:p11);
  \draw[-,black] ([xshift=0.2em]pic cs:p12) -- (pic cs:p12);
  \draw[-,black] ([xshift=0.2em,yshift=0.8em]pic cs:p21) -- ([yshift=0.8em]pic cs:p21);
  \draw[-,black] ([xshift=0.2em]pic cs:p22) -- (pic cs:p22);
  \draw[-,black] ([xshift=0.2em,yshift=0.8em]pic cs:p31) -- ([yshift=0.8em]pic cs:p31);
  \draw[-,black] ([xshift=0.2em]pic cs:p32) -- (pic cs:p32);
  \draw[-,black] ([xshift=0.2em,yshift=0.8em]pic cs:p52) -- ([yshift=0.8em]pic cs:p52);
  \draw[-,black] ([xshift=0.2em]pic cs:p51) -- (pic cs:p51);
  \draw[-,black] ([xshift=0.2em,yshift=0.8em]pic cs:p62) -- ([yshift=0.8em]pic cs:p62);
  \draw[-,black] ([xshift=0.2em]pic cs:p61) -- (pic cs:p61);
\end{tikzpicture}

\label{table:family}

\end{table}

\section{Numerical solutions: a few examples}
\label{Sec:Numerical}

In this section, we present numerical solutions to the static,
spherically symmetric, semiclassical real EKG system described by
Eqs.~(\ref{Eq:KGStaticSphSym}),~(\ref{Eq:EinsteinSphSym1}), and~(\ref{Eq:EinsteinSphSym2}). These
solutions complement the mathematical analysis of the previous
section.  Specifically, we obtain three particular solutions as
representative examples of the three types of solutions that have not
been presented so far in the literature (see
Table~\ref{table:family}), that is, a multi-$\ell$ boson star, a
multi-state $\ell$-boson star, and a multi-$\ell$ multi-state boson
star.

To proceed, from this point onward, in addition to $\hbar=c=1$, we also
set $G=1$, such that all quantities are dimensionless and measured in
Planck units, although the solutions can be rescaled arbitrarily using
a symmetry transformation, as we explain later.  In practice, we solve
the semiclassical real EKG
system~(\ref{Eq:KGStaticSphSym}),~(\ref{Eq:EinsteinSphSym1}), and~(\ref{Eq:EinsteinSphSym2})
expressed in the form
\begin{subequations}\label{eqs.num}
\begin{eqnarray}
\psi_{n\ell}''&=& -\left[\gamma^{2}+1-(2\ell+1)r^{2}\gamma^{2} \left( \frac{\ell(\ell+1)}{r^2} + {m_0}^2\right){(\psi_{n\ell})}^{2} \right]\frac{\psi_{n\ell}'}{r}
- \left(\frac{{(\omega_{n\ell})}^{2}}{\alpha^{2}} -\frac{\ell(\ell+1)}{r^2}-{m_0}^2\right) \gamma^{2}\psi_{n\ell}  , \label{eqs.num1} \\
\gamma' &=&\sum_{n\ell}\frac{2\ell+1}{2}r\gamma\left[ \left( \frac{{(\omega_{n\ell})}^2}{\alpha^2}+\frac{\ell(\ell+1)}{r^2}
+ {m_0}^2\right)\gamma^2{(\psi_{n\ell})}^2+{(\psi_{n\ell}')}^2\right]-\left(\frac{\gamma^2-1}{2r}\right)\gamma ,\label{eqs.num2} \\
\alpha' &=&\sum_{n\ell}\frac{2\ell+1}{2}r\alpha\left[ \left( \frac{{(\omega_{n\ell})}^2}{\alpha^2}-\frac{\ell(\ell+1)}{r^2}
-{m_0}^2\right)\gamma^2{(\psi_{n\ell})}^2+{(\psi_{n\ell}')}^2\right]+\left(\frac{\gamma^2-1}{2r}\right)\alpha , \label{eqs.num3}
\end{eqnarray}
\end{subequations}
where for convenience we have introduced the rescaled fields
\begin{equation}
 \psi_{n\ell} = \sqrt{\frac{N_{n\ell m}}{\omega_{n\ell}}}v_{n\ell} .
\end{equation}
Finally, one can choose an arbitrary value for the mass $m_0$, since
solutions for a different value can then be obtained by a simple
rescaling; see Eq.~(\ref{eq.transformation}) below.  In particular, we
set $m_0=1$ for the numerical integrations, but present the results in
an $m_0$-independent form.  Notice that from the normalization
condition in Eq.~(\ref{Eq:normalization}) one can read off the number
of particles in the different states using expression
\begin{equation}\label{eq.oc.numbers}
 N_{n\ell m} = \omega_{n\ell}\int_0^{\infty}{(\psi_{n\ell})}^2 \frac{\gamma}{\alpha} r^2 dr .
\end{equation}
The choice of appropriate boundary conditions must guarantee that the
boson star solutions are regular and asymptotically flat, and
additionally that they possess finite total energy and finite energy
density everywhere.  Demanding regularity at the origin, i.e.,
\begin{subequations}
\label{eq.conditions1}
\begin{eqnarray}
\psi_{n\ell}(r) &=& \frac{\psi_{n\ell}^0}{2\ell+1} \: r^{\ell}  , \\
\psi_{n\ell}'(r) &=& \frac{\ell\psi_{n\ell}^0}{2\ell+1} \: r^{\ell-1}  , \\
\alpha(r) &=& 1  ,\\
\gamma(r) &=& 1 ,
\end{eqnarray}
\end{subequations}
when $r\to 0$, and a vanishing field at infinity,
one obtains a nonlinear multiple-eigenvalue problem for the different
mode-frequencies $\omega_{n\ell}$.  Here, $\psi_{n\ell}^0$ are some
arbitrary positive constants related to the number of particles in the
different energy levels, and with no loss of generality, we have fixed
the value of the lapse function at the origin to 1. Notice that,
since the system of equations is invariant under $(\alpha,
\omega_{n\ell})\mapsto\lambda (\alpha, \omega_{n\ell})$, with some
positive arbitrary constant $\lambda$, one can always rescale the
value of the lapse function in such a way that $\alpha(r\to\infty)=1$,
as we do later.

The integration of the system is performed numerically using a
shooting algorithm to find the frequencies $\omega_{n\ell}$. To
proceed, one integrates the system of Eqs.~(\ref{eqs.num}) outward,
starting from the initial conditions in Eqs.~(\ref{eq.conditions1}) at
a point very close to the origin (we used $r_0=5\times 10^{-6}$), and
search for the values of the frequencies $\omega_{n\ell}$ to match the
asymptotic behavior of the mode functions until the shooting parameter
converges to the desired accuracy.
As we already mentioned, for simplicity we have assumed $m_0=1$,
although the solutions can be rescaled to an arbitrary value of the
mass parameter using the invariance of the system under the
transformation
\begin{equation}\label{eq.transformation}
 m_0 \mapsto \lambda m_0,\quad \omega_{n\ell}\mapsto \lambda \omega_{n\ell}, \quad r\mapsto \lambda^{-1} r,
\end{equation}
with the functions $\psi_{n\ell}$, $\alpha$ and $\gamma$ unchanged.
Under this transformation, the occupation
numbers~(\ref{eq.oc.numbers}) change according to $N_{n\ell m}\mapsto
\lambda^{-2}N_{n\ell m}$.  Note that, for instance, $\lambda\sim
10^{-50}$ in the case of an ultralight axion dark matter particle of
mass $m_0\sim 10^{-22}\,$eV.

\begin{figure}[h]
    \subfloat[Multi-$\ell$ boson star \label{figa}]{\includegraphics[angle=0,width=0.3\textwidth]{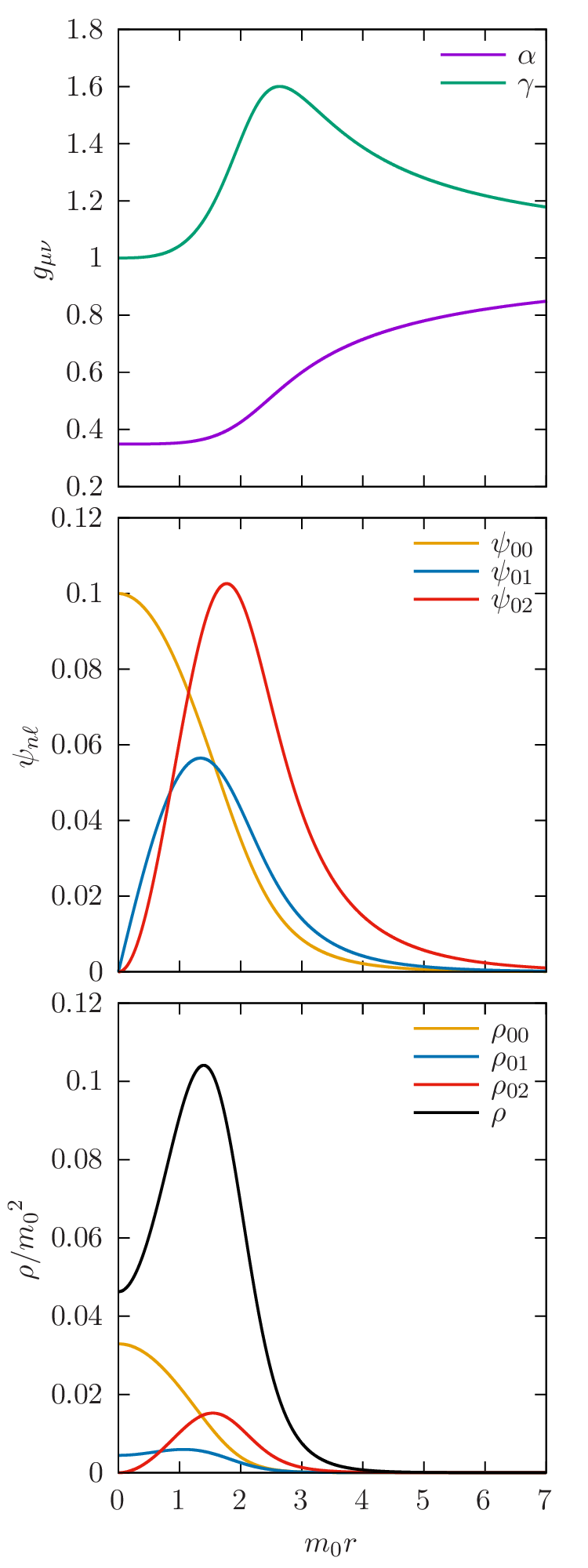}}
    \subfloat[Multi-state $\ell$-boson star \label{figb}]{\includegraphics[angle=0,width=0.3\textwidth]{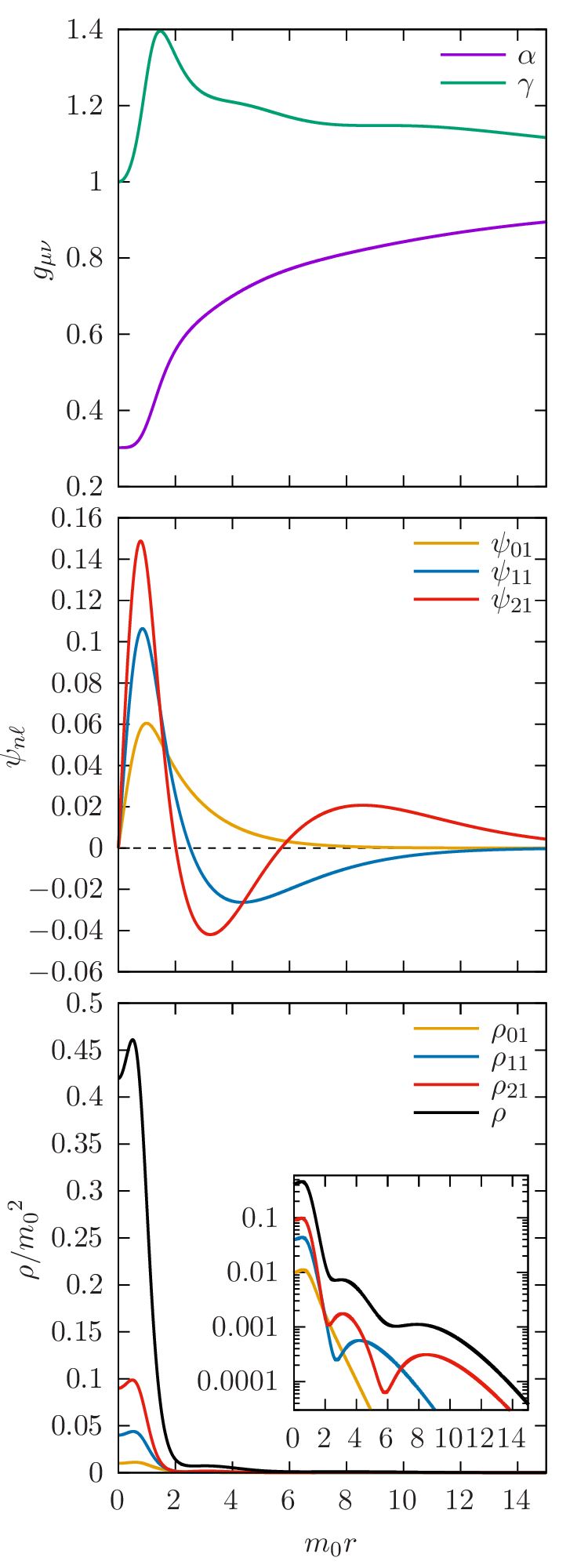}}
    \subfloat[Multi-$\ell$ multi-state boson star \label{figc}]{\includegraphics[angle=0,width=0.3\textwidth]{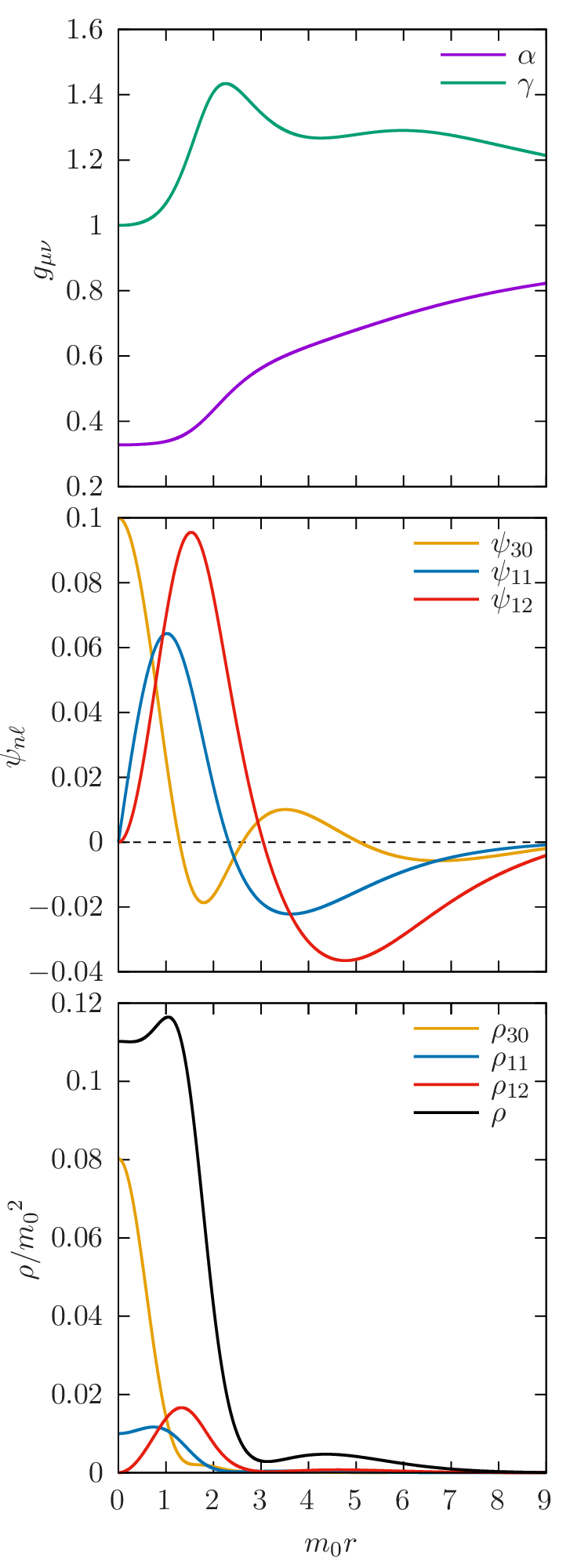}}
    \caption{As examples we describe here three particular solutions,
      each one of a different type, as indicated in the
      subfigures. For all cases we show the metric components,
      $\alpha(r)$ and $\gamma(r)$ (top panels); the state components
      $\psi_{n \ell}(r)$ (middle panels); and the energy density
      $\rho(r)$, as well as the individual components
      $\rho_{n\ell}(r)$ (bottom panels). Note that $\rho$ can be
      obtained simply by summing its components (including all
      degenerate states in $m$).  Thus, we have
      $\rho=\rho_{00}+3\rho_{01}+5\rho_{02}$ in~\protect\subref{figa},
      $\rho=3\rho_{01}+3\rho_{11}+3\rho_{21}$
      in~\protect\subref{figb}, and
      $\rho=\rho_{30}+3\rho_{11}+5\rho_{12}$
      in~\protect\subref{figc}. We show more properties of these
      solutions in Table~\ref{table:plots}. Finally, we note that the actual numerical integration domain in $r$ is much larger than what is shown (for clarity) in the figures.}
    \label{figall}
\end{figure}

In Figure~\ref{figall}, we present the results of our numerical
solutions.  In the left column ---subfigures~\subref{figa}--- we
present a multi-$\ell$ boson star solution. The configuration
displayed is characterized by the quantum numbers $n=0$ and
$\ell=0,\,1,\,2$.\footnote{Note that, as in our previous works, 
$n$ refers to the number of nodes of the radial function in the interval $0 < r < \infty$. Hence, it should be compared to the ``radial quantum number" in the theory of the hydrogen atom rather than the ``principal quantum number".}
Unlike the $\ell$-boson stars
introduced in~\cite{Alcubierre:2018ahf}, these new solutions do not
require all the individual fields to have the same amplitude, and,
still, the resulting spacetime is spherically symmetric. In the
central column ---subfigures~\subref{figb}--- we present a multi-state
$\ell$-boson star solution. The configuration displayed is
characterized by the quantum numbers $n=0,\,1,\,2$ and $\ell=1$.  In
the right column ---subfigures~\subref{figc}--- we present a
multi-$\ell$ multi-state boson star solution. The configuration
displayed is characterized by the quantum numbers $n=1,\,3$ and
$\ell=0,\,1,\,2$.  In all three cases the lapse function $\alpha$ and
the metric component $\gamma$ of each configuration are shown in the
top panels, the radial profiles for the different wave functions
$\psi_{n\ell}$ that are excited in the configuration are shown in the
middle panels, and the total energy density, as well as the density of
the individual energy levels are shown in the bottom panel.  In
Table~\ref{table:plots}, we report the main numbers associated with
these configurations.

\begin{table}[t]\centering
\caption{Quantum numbers, amplitudes, eigenfrequencies, particle
  numbers in each state, and total particle number for the
  configurations presented in Fig.~\ref{figall}.  Note that for the
  case of an ultralight axion $m_0\sim 10^{-50}$ in natural units;
  hence in this example the particle numbers are large for the
  configurations presented.}\label{table:plots}
\begin{tabular}{l c c c c c c c}
  \hline
  \hline
Name                                & $n$ & $\ell$ & $\psi^0_{n\ell}/{m_0}^\ell$ & $\omega_{n\ell}/m_0$ & ${m_0}^2 N_{nl}$ & ${m_0}^2 N$ & Fig.                        \\
  \hline
                                    & $0$ & $0$    & $0.1$                       & $0.5278$              & $0.0195$         &             &                             \\
Multi-$\ell$ boson star             & $0$ & $1$    & $0.2$                       & $0.6453$              & $0.0243$         & 0.8134      & \ref{figall}\subref{figa} \\
                                    & $0$ & $2$    & $0.4$                       & $0.7736$              & $0.1442$         &             &                             \\
\hline
                                    & $0$ & $1$    & $0.3$                       & $0.7438$              & $0.0289$         &             &                             \\
Multi-state $\ell$-boson star       & $1$ & $1$    & $0.6$                       & $0.8235$              & $ 0.1150$        & 1.3884      & \ref{figall}\subref{figb} \\
                                    & $2$ & $1$    & $0.9$                       & $0.8792$              & $ 0.3189$        &             &                             \\
\hline
                                    & $3$ & $0$    & $0.1$                       & $0.8679$              & $0.0133$         &             &                             \\
Multi-$\ell$ multi-state boson star & $1$ & $1$    & $0.3$                       & $0.7497$              & $0.0439$         & 1.2745      & \ref{figall}\subref{figc} \\
                                    & $1$ & $2$    & $0.5$                       & $0.8247$              & $0.2259$         &             &                             \\ 
  \hline
  \hline
\end{tabular}
\end{table}

In this section we presented just three particular solutions as
illustrative examples of the three new types of boson stars described
in this work. However, going into further analysis of such solutions
is beyond the scope of this article. Instead, we expect to carry out a
more detailed study in a separate work.

\section{Discussion and conclusions}
\label{Sec:Discussion}

We have shown that static, spherically symmetric boson star
configurations~\cite{Kaup68,Ruffini:1969qy,Colpi:1986ye,Friedberg87,Gleiser:1988rq,Lee:1988av,Guzman:2004wj,Bizon:2000,Chavanis:2011cz,Jetzer:1991jr,Schunck:2003kk,Liebling:2012fv,Zhang:2018slz,Visinelli:2021uve}
and many of their generalizations (including $\ell$-boson
stars~\cite{Alcubierre:2018ahf,Alcubierre:2019qnh,Alcubierre:2021mvs,Alcubierre:2021psa,Jaramillo:2022zwg}
and multi-state boson
stars~\cite{Matos2007,Bernal:2009zy,UrenaLopez:2010ur}), arise
naturally within the semiclassical gravity approximation in quantum
field theory.  Furthermore, we have found new possible
generalizations, namely the multi-$\ell$ multi-state boson stars, that
represent the most general solutions to the $N$-particle, static,
spherically symmetric, semiclassical real EKG system, in which the
total number of particles is definite.

Our approach is based on the expansion of a single, real, free quantum scalar
field in terms of a linear combination of creation and annihilation
operators.  We then construct the Hilbert space by successive
applications of creation operators on the vacuum state, that we assume
is well defined (which is guaranteed for the static configurations
considered in this article).  Taking particular Fock states with a
definite number of particles, the expectation value of the
renormalized stress energy-momentum tensor operator takes the same
form as its counterpart in a classical theory with $N$ fields.
Each of these $N$ fields accounts for one excitation mode of the
quantum field, and corresponds to the quantum particles of the
system. The number of particles contained in each mode is then fixed
by the number of classical fields with the same quantum numbers
and can be chosen as a free parameter.  In this way, standard boson
stars correspond to the population of only one mode (the ground state)
of the quantum field, for which $n=\ell=m=0$. Other self-gravitating
scalar configurations such as $\ell$-boson stars and multi-state boson
stars are naturally related to them and are just different manifestations of the quantum
fluctuations of the same scalar field with different modes populated.

Regarding $\ell$-boson stars in particular, we would like to highlight
an important difference between the classical and semiclassical
interpretations: the construction of the classical solutions described
in our previous works might be considered somehow ``artificial'', in
the sense that they must consist of a particular combination of
$2\ell+1$ independent, complex scalar fields having the exact same
radial amplitude in order to constitute as a whole a spherically
symmetric matter distribution.
On the other hand, within the semiclassical gravity approximation, the same $\ell$-boson star configurations are obtained more naturally, starting from a single real quantum scalar field in a state that describes a distribution of particles
populating a given energy level and containing angular momentum.
Apart from providing a more natural explanation for their existence, the semiclassical description of $\ell$-boson stars should have other interesting applications. 
For example, a relevant problem is analyzing the dynamical stability of these configurations in semiclassical gravity. This is clearly a much more difficult problem 
that requires generalizing our framework for spacetimes being perturbed off a static one.

Further, in the realm of an ultralight scalar field/fuzzy dark matter component, the
diversity of self-gravitating structures that might model the dark
matter halos is greatly enhanced in the semiclassical theory. 
Standard $\ell=0$ boson stars have been considered as a possible explanation for the dark matter distribution 
in small galaxies, such as dwarf spheroidals~\cite{Gonzalez-Morales:2016yaf}. However, it is clear that these configurations alone are not sufficient to describe the properties observed in 
galaxies spanning different size scales.
On the one hand, the radius of stable $\ell=0$ boson stars decreases when their total mass increases, in stark contrast to what is being observed in real galaxies.
On the other hand, their mass density decreases exponentially with the radius at large distances, and they
cannot accommodate the flatness of the rotation curves observed in spiral galaxies (although they can still describe
the core of galaxies like the Milky Way~\cite{DeMartino:2018zkx}). 
The multi-$\ell$ multi-state configurations offer the possibility to account for the desired behavior of the dark matter distribution in large galaxies beyond the core region. 
Likewise, several examples of anomalous halo systems recently reported in \cite{Atlas3D}
could potentially be explained by means of scalar field configurations
with nontrivial values of $n$, $\ell$, and $m$. 
The connection between the multi-$\ell$ multi-state boson star solutions and the configurations obtained in numerical 
simulations~\cite{Schive:2014dra,Schive:2014hza,Schwabe:2016rze,Veltmaat:2016rxo,Du:2016aik} is left as an open question.

With regard to dark compact objects, the state of maximal compactness of $\ell$-boson stars has been identified to grow with $\ell$, as discussed in~\cite{Alcubierre:2021psa}.
This could lead to potential observational signatures in the gravitational wave spectra resulting from the merger of these objects~\cite{Palenzuela:2017kcg}.

In this article we have only
presented some representative cases by solving numerically the
spherically symmetric, static semiclassical system of equations for certain quantum numbers; however
a more exhaustive analysis is required in order to compare our
solutions with the astrophysical data.
In conclusion, the
interpretation of boson stars within the semiclassical gravity
approximation considerably increases the possibility for boson fields
to describe more realistic astrophysical systems.


\acknowledgments A.D.T. acknowledges conversations with Cristian
Armendariz-Picon and Daniel Sudarsky. O.S. wishes to thank stimulating discussions with Robert Oeckl, Jorge Pullin, Armando Roque, and Thomas Zannias. This work was supported in part
by the CONACYT Network Projects No. 376127 ``Sombras, lentes y ondas
gravitatorias generadas por objetos compactos astrof\'isicos'',
No. 304001 ``Estudio de campos escalares con aplicaciones en
cosmolog\'ia y astrof\'isica'', No. 1406300 ``Explorando los confines
de las teor\'ias relativistas de la gravitaci\'on y sus
consecuencias'', and No. 286897 ``Materia oscura: Implicaciones de sus
propiedades fundamentales en las observaciones astrof\'isicas y
cosmol\'ogicas'', by DGAPA-UNAM grant IN105920, by the European Union's
Horizon 2020 research and innovation (RISE) program
H2020-MSCA-RISE-2017 Grant No. FunFiCO-777740 and by a CIC grant to
Universidad Michoacana de San Nicol\'as de Hidalgo.

\appendix

\section{Stress energy-momentum tensor}
\label{app.energy-momentum}

In this appendix we summarize the results for the energy
density~(\ref{eq.density.gen}), momentum flux~(\ref{eq.flux.gen}) and
spatial stress tensor~(\ref{eq.stress.gen}) of an arbitrary quantum
state on a static spacetime that are used in
section~\ref{Sec:Static}. These expressions are
\begin{subequations}\label{eqs.general}
\begin{eqnarray}
\rho &=& \sum_{I,J}\frac{1}{4\sqrt{\omega_I\omega_J}} \left\{ \langle\hat{a}_I\hat{a}_J \rangle \left[ 
\left( -\frac{\omega_I\omega_J}{\alpha^2} + {m_0}^2 \right) u_I u_J + (D_k u_I)(D^k u_J) \right]
e^{-i(\omega_I+\omega_J)t}\right. \nonumber \\
&& \hspace{1.9cm} \left. + \langle\hat{a}^{\dagger}_I\hat{a}_J\rangle \left[ 
\left( +\frac{\omega_I\omega_J}{\alpha^2} + {m_0}^2 \right) u_I^* u_J
+ (D_k u_I^*)(D^k u_J) \right] e^{+i(\omega_I-\omega_J)t} + c.c.\right\} ,
\label{Eq:rho} \\
 j_k &=& \sum_{I,J}\frac{1}{4\sqrt{\omega_I\omega_J}} \frac{i}{\alpha}
 \left\{ \langle\hat{a}_I\hat{a}_J \rangle 
 \left[ -\omega_J (D_k u_I) u_J - \omega_I u_I (D_k u_J) \right]
e^{-i(\omega_I+\omega_J)t} \right. \nonumber \\
&& \hspace{2.25cm}\left.+ \langle\hat{a}^{\dagger}_I\hat{a}_J \rangle 
\left[ - \omega_J (D_k u_I^*) u_J + \omega_I u_I^* (D_k u_J) \right]
 e^{+i(\omega_I-\omega_J)t} - c.c.\right\} ,
\label{Eq:j}
\end{eqnarray}
and
\begin{eqnarray}
S_{ij} &=& \sum_{I,J}\frac{1}{4\sqrt{\omega_I\omega_J}} \nonumber \\
 &\times& \left\{ \langle\hat{a}_I\hat{a}_J\rangle \left[ (D_i u_I)(D_j u_J) + (D_j u_I)(D_i u_J) - \gamma_{ij}\left(
 \left(+ \frac{\omega_I\omega_J}{\alpha^2} + {m_0}^2 \right) u_I u_J + (D_k u_I)(D^k u_J) \right)\right]e^{-i(\omega_I+\omega_J)t}\right. \nonumber \\
&& \left.\hspace{-0.06cm}+ \langle\hat{a}^{\dagger}_I\hat{a}_J\rangle 
\left[ (D_i u_I^*)(D_j u_J) + (D_j u_I^*)(D_i u_J) - \gamma_{ij}\left(
 \left( -\frac{\omega_I\omega_J}{\alpha^2} + {m_0}^2 \right) u_I^* u_J + (D_k u_I^*)(D^k u_J) \right)\right]e^{+i(\omega_I-\omega_J)t} \right. \nonumber\\
&& \left. \hspace{-0.06cm}+ c.c.\right\} ,
\label{Eq:Sij}
\end{eqnarray}
\end{subequations}
where $c.c.$ stands for complex conjugation.

In this article, we are interested in {\it static configurations},
that is in states satisfying the property that $\rho$ and $S_{ij}$ are
time-independent and $j_k = 0$, such that the expectation value of the
stress energy-momentum tensor is compatible, through the semiclassical
Einstein equations, with the static spacetime
metric~(\ref{Eq:StaticMetric}).  It is clear from
Eqs.~(\ref{eqs.general}) that $\rho$, $j_k$, and $S_{ij}$ are
time-independent if $\langle\hat{a}_I\hat{a}_J\rangle = 0$ for all
$I,J$ and $\langle\hat{a}_I^\dagger\hat{a}_J\rangle = 0$ whenever
$\omega_I\neq \omega_J$. Furthermore, assuming the convention in
footnote~\ref{Footnote:Five}, the additional condition $\langle
\hat{N}_I \rangle = \langle \hat{N}_{I'} \rangle$ implies that $j_k =
0$.

The question of whether these conditions are also necessary for a
static configuration is more subtle. To proceed, we introduce the
notion of {\it stationary states}, defined as states in the
Schr\"odinger picture which are invariant under time translations. In
quantum mechanics time translations are represented by unitary
operators of the form \mbox{$\hat{\mathcal{U}}[t_0,t] =
\exp[-i\hat{H}(t-t_0)]$}, where the Hamiltonian $\hat{H}$ is the
generator of the transformations.  If a state $|\psi\rangle$ is
invariant under time evolution, then $\hat{\mathcal{U}}[t_0,t]
|\psi\rangle = |\psi\rangle$, apart from a possible phase that does
not contribute to physical observables. Hence, stationary states are
eigenvectors of the Hamiltonian operator. If, in addition, the
condition $\langle \hat{N}_I \rangle = \langle \hat{N}_{I'} \rangle$
is satisfied, the state is called {\it static}.

As we show now, static states give rise to static configurations. To
prove this, we first observe that the Hamiltonian can be defined as
\begin{equation}
\hat{H} := \int_{\Sigma_t} \hat{T}_{\mu\nu} k^\mu n^\nu d\gamma ,
\end{equation}
with $n_\nu$ and $d\gamma$ given as in Eq.~(\ref{scalar}), and $k =
\partial_t$ the timelike Killing vector field associated with the
static symmetry.  Taking $\Sigma_t$ to be a $t=const.$ hypersurface,
such that $k^\nu = \alpha n^\nu$, and using
Eqs.~(\ref{energy.momentum.2}), (\ref{Eq:StaticMetric}),
(\ref{Eq:fIStatic}), (\ref{Eq:KGStatic}), (\ref{Eq:ScalarProd}), and~(\ref{Eq:uIOrtho}) and integration by parts, we obtain
\begin{equation}
\hat{H} = \sum\limits_I\hat{N}_I \omega_I .
\end{equation}
In particular, it follows that each of the states $\vert
N_1,N_2,\ldots \rangle$ is an eigenfunction of $\hat{H}$ with energy
$E = \sum_I N_I\omega_I$. Any stationary state can be written as a
superposition of eigenfunctions of the form $\vert N_1,N_2,\ldots
\rangle$ which have the same definite value of the energy $E$ (note
that if $E$ is degenerate, this superposition might contain several
such eigenfunctions.) Since $\omega_I > 0$ for each $I$, it follows
that $\hat{a}_I \hat{a}_J\vert N_1,N_2,\ldots \rangle$ is either zero,
or is an eigenfunction of $\hat{H}$ with energy smaller than $E$. This
implies that $\langle\hat{a}_I\hat{a}_J\rangle = 0$ for all
$I,J$. Similarly, $\hat{a}_I\vert N_1,N_2,\ldots \rangle$ is either
zero, or is an eigenfunction of $\hat{H}$ with energy $E - \omega_I$,
which implies that $\langle\hat{a}_I^\dagger\hat{a}_J\rangle = 0$ if
$\omega_I\neq \omega_J$.  These conditions imply that $\rho$, $j_k$,
and $S_{ij}$ are time-independent. Therefore, if in addition the state
is static, we conclude that the configuration is static.

An interesting question is whether or not static configurations can
only be sourced by static states. For a real scalar field, we have not
been able to obtain a particular realization of a non-static state
that is associated with a static configuration. However, in the next
section we provide such a counterexample in the complex scalar field
theory.

 \section{Complex scalar fields}
 \label{sec.complex}

In this appendix, we generalize the static, spherically symmetric,
semiclassical real EKG system to the case of a complex scalar
field. The main interest of this extension is that, as we are going to
demonstrate below (and in stark contrast to the real case), when the
field is complex there are coherent states which are compatible with a
static spacetime background.

The construction of the semiclassical, complex EKG system is very
similar to that presented in the main text, so for the sake of
simplicity we will only stress the main differences with respect to
the real case. When expressed in terms of the creation
$(\hat{a}_I^{\dagger}$, $\hat{b}_I^{\dagger})$ and annihilation
$(\hat{a}_I$, $\hat{b}_I)$ operators, a complex scalar field takes the
form
\begin{equation}
\hat{\phi}(x) = \sum_I\left[ \hat{a}_I f_I(x) +  \hat{b}^{\dagger}_I f^*_I(x) \right],
\label{eq.complex}
\end{equation}
where $f_1,f_2,\ldots$ is an orthonormal set of basis functions of the
subspace $X_+$, and the creation and annihilation operators satisfy
the standard commutation relations
$[\hat{a}_I,\hat{a}_J^{\dagger}]=[\hat{b}_I,\hat{b}_J^{\dagger}]=\delta_{IJ}$,
with all other possible commutators vanishing. The Hilbert space
$\mathscr{H}$ can be constructed by successive application of creation
operators on the vacuum state. The elements of this construction
$|N^a_1,\ldots,N_K^a,\ldots,N^b_1,\ldots,N_K^b,\ldots\rangle$ are the
eigenstates of the particle and antiparticle number operators $\hat{N}_I^a=\hat{a}_I^{\dagger}\hat{a}_I$ and $\hat{N}_I^b=\hat{b}_I^{\dagger} \hat{b}_I$, respectively.
As in the real case, the coherent
states are defined as the eigenstates of the annihilation operators
$\hat{a}_I$ and $\hat{b}_I$.  Note that the main difference with
respect to the real field is that the complex theory contains
particles and antiparticles, although they are indistinguishable
unless we introduce interactions with other fields that break the
degeneracy, which is not the case here.
 
The stress  energy-momentum tensor operator associated  with a complex
scalar field takes the form
 \begin{equation}
\hat{T}_{\mu\nu} = 
(\nabla_{\mu}\hat{\phi})(\nabla_{\nu}\hat{\phi})^{\dagger} + (\nabla_{\nu}\hat{\phi})(\nabla_{\mu}\hat{\phi})^{\dagger}
- g_{\mu\nu}\left[ (\nabla_{\alpha}\hat{\phi})(\nabla^{\alpha}\hat{\phi})^{\dagger}
 + {m_0}^2\hat{\phi}\hat{\phi}^{\dagger} \right]  .
\label{energy.momentum.complex}
 \end{equation}
Introducing the decomposition~(\ref{eq.complex})
into~(\ref{energy.momentum.complex}), we obtain
\begin{equation}
\hat{T}_{\mu\nu} = 
\sum_{I,J}\left[\hat{a}_I\hat{a}^{\dagger}_J T_{\mu\nu}(f_I,f_J^*) 
 + \hat{a}_I\hat{b}_J T_{\mu\nu}(f_I,f_J) + \hat{b}_I^{\dagger}\hat{a}^{\dagger}_J T_{\mu\nu}(f_I^*,f_J^*) 
 + \hat{b}_I^{\dagger}\hat{b}_J T_{\mu\nu}(f_I^*,f_J) \right] ,
\label{energy.momentum.complex2}
\end{equation}
with $T_{\mu\nu}(f_I,f_J)$ as in Eq.~(\ref{eq.T.f's}).

If the spacetime is static, the energy density~(\ref{eq.density.gen}),
momentum flux~(\ref{eq.flux.gen}), and spatial stress
tensor~(\ref{eq.stress.gen}) take the form
\begin{subequations}
\label{eqs.general.complex}
\begin{eqnarray}
 \rho &=& \sum_{I,J}\frac{1}{2\sqrt{\omega_I\omega_J}} \left\{ \langle\hat{a}_I^{\dagger}\hat{a}_J\rangle \left[ 
\left( + \frac{\omega_I\omega_J}{\alpha^2} + {m_0}^2 \right) u_I^* u_J + (D_k u_I)(D^k u^*_J) \right]
e^{+i(\omega_I-\omega_J)t}\right. \nonumber \\
&& \hspace{1.95cm}\left.+ \langle\hat{a}_I\hat{b}_J\rangle \left[ 
\left( - \frac{\omega_I\omega_J}{\alpha^2} + {m_0}^2 \right) u_I u_J + (D_k u_I)(D^k u_J) \right] e^{-i(\omega_I+\omega_J)t} \right. \nonumber \\
&& \hspace{1.95cm}\left.+ \langle\hat{b}^{\dagger}_I\hat{a}_J^{\dagger}\rangle \left[ 
\left( - \frac{\omega_I\omega_J}{\alpha^2} + {m_0}^2 \right) u_I^* u^*_J + (D_k u_I^*)(D^k u^*_J) \right] e^{+i(\omega_I+\omega_J)t} \right. \nonumber \\
&& \hspace{1.95cm}\left.+ \langle\hat{b}^{\dagger}_I\hat{b}_J\rangle \left[ 
\left( + \frac{\omega_I\omega_J}{\alpha^2} + {m_0}^2 \right) u_I^* u_J + (D_k u_I^*)(D^k u_J) \right] e^{+i(\omega_I-\omega_J)t} \right\}  , \\
j_k &=& \sum_{I,J}\frac{1}{2\sqrt{\omega_I\omega_J}} \frac{i}{\alpha}
 \left\{ \langle\hat{a}^{\dagger}_I\hat{a}_J\rangle 
 \left[ -\omega_J (D_k u_I^*) u_J  + \omega_I u_I^* (D_k u_J) \right]
e^{-i(\omega_I-\omega_J)t}\right.\nonumber\\
&& \hspace{2.25cm}\left.+ \langle\hat{a}_I\hat{b}_J\rangle 
\left[ -\omega_J (D_ku_I) u_J - \omega_I u_I (D_k u_J) \right]
 e^{-i(\omega_I+\omega_J)t} \right.\nonumber \\
 && \hspace{2.25cm}\left.+ \langle\hat{b}^{\dagger}_I\hat{a}_J^{\dagger}\rangle
\left[ +\omega_J (D_k u_I^*) u_J^* + \omega_I u_I^* (D_k u_J^*) \right]
 e^{+i(\omega_I+\omega_J)t} \right.\nonumber\\
 && \hspace{2.25cm}\left.+ \langle\hat{b}^{\dagger}_I\hat{b}_J\rangle
\left[ -\omega_J (D_k u_I^*) u_J  + \omega_I u_I^* (D_k u_J) \right]
 e^{+i(\omega_I-\omega_J)t} \right\} ,
\end{eqnarray}
 and
\begin{eqnarray}
S_{ij} &=& \sum_{I,J}\frac{1}{2\sqrt{\omega_I\omega_J}} \\
&\times& \left\{ \langle\hat{a}^{\dagger}_I\hat{a}_J\rangle \left[ (D_i u_I^*)(D_j u_J) + (D_j u_I^*)(D_i u_J) - \gamma_{ij}\left(
 \left(- \frac{\omega_I\omega_J}{\alpha^2} + {m_0}^2 \right) u_I^* u_J + (D_k u_I^*)(D^k u_J) \right)\right]e^{+i(\omega_I-\omega_J)t}\right.\nonumber\\
&& \left.\hspace{-0.04cm}+ \langle\hat{a}_I\hat{b}_J\rangle 
\left[ (D_i u_I)(D_j u_J) + (D_j u_I)(D_i u_J) - \gamma_{ij}\left(
 \left( +\frac{\omega_I\omega_J}{\alpha^2} + {m_0}^2 \right) u_I u_J + (D_k u_I)(D^k u_J) \right)\right]e^{-i(\omega_I+\omega_J)t} \right. \nonumber\\
&& \left.\hspace{-0.04cm}+ \langle\hat{b}^{\dagger}_I\hat{a}_J^{\dagger}\rangle 
\left[ (D_i u_I^*)(D_j u_J^*) + (D_j u_I^*)(D_i u_J^*) - \gamma_{ij}\left(
 \left( +\frac{\omega_I\omega_J}{\alpha^2} + {m_0}^2 \right) u_I^* u_J^* + (D_k u_I^*)(D^k u_J^*) \right)\right]e^{+i(\omega_I+\omega_J)t} \right. \nonumber\\
&& \left.\hspace{-0.04cm}+ \langle\hat{b}^{\dagger}_I\hat{b}_J\rangle 
\left[ (D_i u_I^*)(D_j u_J) + (D_j u_I^*)(D_i u_J) - \gamma_{ij} \left(
\left( +\frac{\omega_I\omega_J}{\alpha^2} + {m_0}^2 \right) u_I^* u_J + (D_k u_I^*)(D^k u_J) \right)\right]e^{-i(\omega_I-\omega_J)t} \right\} , \nonumber
\end{eqnarray}
\end{subequations}
where we have used Wick's ordering to obtain these expressions.
Comparing them with those of Eq.~(\ref{eqs.general}) we find that,
apart from an overall factor of 2, they differ in some of the
expectation values which are quadratic in creation and annihilation
operators. Of course, if we choose a state with a definite number of
particles, for which $\langle\hat{a}^{\dagger}_I\hat{a}_J\rangle=
N_I\delta_{IJ}$ and $\langle\hat{a}_I\hat{b}_J\rangle =
\langle\hat{a}^{\dagger}_I\hat{b}^{\dagger}_J\rangle =
\langle\hat{b}^{\dagger}_I\hat{b}_J\rangle=0$,
Eqs.~(\ref{eqs.general}) and~(\ref{eqs.general.complex}) coincide, and
we recover the relations in Eqs.~(\ref{eq.T.coherent.state}).
The same holds true for a state with a definite number of
antiparticles.  Furthermore, it is even possible to have a state with
a definite number of particles {\it and} antiparticles simultaneously;
in that case, we only have to replace $N_I$ by $N_I^a+N_I^b$ in
Eqs.~(\ref{eq.T.coherent.state}).  This is because the stress
energy-momentum tensor sources gravity, and this tensor does not
differentiate between particles and antiparticles.  However, if the
field is complex, it is still possible to guarantee a static source
even if the system is not in an eigenstate of the Hamiltonian
operator. This is what happens for coherent states that involve a
single mode $I$ of the particle or antiparticle sector.  For these
states
$\langle\hat{a}^{\dagger}_I\hat{a}_J\rangle=|\alpha_I|^2\delta_{IJ}$,
$\langle\hat{a}_I\hat{b}_J\rangle=\langle\hat{b}^{\dagger}_I\hat{a}_J^{\dagger}\rangle=\langle\hat{b}^{\dagger}_I\hat{b}_J\rangle=0$
(where in this case we have assumed that we only have particles), and
it is easy to convince oneself that we recover
Eqs.~(\ref{eq.T.coherent.state}) with $N_I$ replaced by $|\alpha_I|^2$
and no sum over $I$.  Note that this is not possible in the real case,
where the coefficients
$\langle\hat{a}_I\hat{a}_J\rangle=(\alpha_i)^2\delta_{IJ}$ and
$\langle\hat{a}^{\dagger}_I\hat{a}_J^{\dagger}\rangle=(\alpha_I^*)^2\delta_{IJ}$
of a coherent state
give rise to a time dependency of the source terms that is
not compatible with a static metric.  The main difference of these
solutions with respect to those with a definite number of particles is
that the expectation value of the scalar field evolves nontrivially in
time.



\begin{thebibliography}{99}

\bibitem{Kaup68} D.J.~Kaup, ``Klein-Gordon geon,'' 
Phys. Rev. {\bf 172}, 1331 (1968)

\bibitem{Ruffini:1969qy}
R.~Ruffini and S.~Bonazzola, ``Systems of selfgravitating particles in general relativity and the concept of an equation of state,'' 
Phys. Rev. {\bf 187}, 1767-1783 (1969)

\bibitem{Colpi:1986ye} M.~Colpi, S.L.~Shapiro and I.~Wasserman, ``Boson stars: Gravitational equilibria of selfinteracting scalar fields,''
Phys. Rev. Lett. {\bf 57}, 2485-2488 (1986)

\bibitem{Friedberg87} R.~Friedberg, T.D.~Lee and Y.~Pang, ``Mini-soliton stars,'' 
Phys. Rev. D {\bf 35}, 3640-3657 (1987)

\bibitem{Gleiser:1988rq} M.~Gleiser, ``Stability of boson stars,''
Phys. Rev. D {\bf 38}, 2376 (1988)

\bibitem{Lee:1988av}
T.D.~Lee and Y.~Pang, ``Stability of mini-boson stars,''
Nucl. Phys. B {\bf 315}, 477 (1989)

\bibitem{Bizon:2000}
P.~Bizon and A. Wasserman, ``On existence of mini-boson stars,"
Commun. Math. Phys. {\bf 215}, 357 (2000) [arXiv:gr-qc/0002034]

\bibitem{Guzman:2004wj} F.S.~Guzman and L.A.~Ure\~na-L\'opez, ``Evolution of the Schr\"odinger-Newton system for a selfgravitating scalar field,''
Phys. Rev. D {\bf 69}, 124033 (2004) [arXiv:gr-qc/0404014]

\bibitem{Chavanis:2011cz} P.H.~Chavanis and T.~Harko, ``Bose-Einstein condensate general relativistic stars,''
Phys. Rev. D {\bf 86}, 064011 (2012) [arXiv:1108.3986 [astro-ph.SR]

\bibitem{Jetzer:1991jr} P.~Jetzer, ``Boson stars,''
Phys. Rept. {\bf 220}, 163-227 (1992)

\bibitem{Schunck:2003kk}
F.E.~Schunck and E.W.~Mielke, ``General relativistic boson stars,'' 
Class. Quant. Grav. {\bf 20}, R301-R356 (2003) [arXiv:0801.0307 [astro-ph]]

\bibitem{Liebling:2012fv}
S.L.~Liebling and C.~Palenzuela, ``Dynamical boson stars,'' 
Living Rev. Rel. {\bf 15}, 6 (2012) [arXiv:1202.5809 [gr-qc]]

\bibitem{Zhang:2018slz}
E.~Braaten and H.~Zhang, ``Axion stars,'' 
Symmetry {\bf 12}, 25 (2019) [arXiv:1810.11473]

\bibitem{Visinelli:2021uve}
L.~Visinelli, ``Boson stars and oscillatons: A review,'' 
Int. J. Mod. Phys. D {\bf 30}, 2130006 (2021) [arXiv:2109.05481 [gr-qc]]

\bibitem{Lee:1991ax} T.D.~Lee and Y.~Pang, ``Nontopological solitons,''
Phys. Rept. {\bf 221}, 251-350 (1992)

\bibitem{Rajaraman} 
R. Rajaraman, {\it Solitons and Instantons:  An Introduction to Solitons and Instantons in Quantum Field Theory,}  
North-Holland (1987)

\bibitem{Manton2007} 
N.~Manton and P.~Sutcliffe, {\it Topological Solitons,}  
Cambridge Monographs on Mathematical Physics. Cambridge University Press (2007) 508pp

\bibitem{EJWeinberg} 
E.J. Weinberg, {\it Classical Solutions in Quantum Field Theory:  Solitons and Instantons in High Energy Physics,}  
Cambridge Monographs on Mathematical Physics. Cambridge University Press (2012) 342pp

\bibitem{Shnir}
Y.M. Shnir, {\it Topological and Non-Topological Solitons in Scalar Field Theories,}
Cambridge Monographs on Mathematical Physics. Cambridge University Press (2018) 278pp

\bibitem{Hogan:1988mp} C.J.~Hogan and M.J.~Rees, ``Axion miniclusters,''
Phys. Lett. B {\bf 205}, 228-230 (1988)

\bibitem{Kolb:1993zz} E.W.~Kolb and I.I.~Tkachev, ``Axion miniclusters and Bose stars,''
Phys. Rev. Lett. {\bf 71}, 3051-3054 (1993) [arXiv:9303313 [hep-ph]]

\bibitem{Torres:2000dw} D.F.~Torres, S.~Capozziello and G.~Lambiase, ``A Supermassive scalar star at the galactic center?,''
Phys. Rev. D {\bf 62}, 104012 (2000) [arXiv:104012 [astro-ph]]

\bibitem{Guzman:2005bs}
F.G.~Guzman, ``Accretion disc onto boson stars: A Way to supplant black holes candidates,''
Phys. Rev. D {\bf 73}, 021501(R) (2006) [arXiv:0512081 [gr-qc]]

\bibitem{Palenzuela:2017kcg}
C.~Palenzuela, P.~Pani, M.~Bezares, V.~Cardoso, L.~Lehner and S.~Liebling, ``Gravitational wave signatures of highly compact boson star binaries,''
Phys. Rev. D {\bf 96}, 104058 (2017) [arXiv:1710.09432 [gr-qc]]

\bibitem{Cardoso:2019rvt}
V.~Cardoso and P.~Pani, ``Testing the nature of dark compact objects: a status report,''
Living Rev. Rel. {\bf 22}, 4 (2019) [arXiv:1904.05363 [gr-qc]]

\bibitem{Sin:1992bg} S-J. Sin, ``Late time cosmological phase transition and galactic halo as Bose liquid,''
Phys. Rev. D {\bf 50}, 3650-3654 (1994) [arXiv:9205208 [hep-ph]]

\bibitem{Schive:2014dra}
H-Y.~Schive, T.~Chiueh and T.~Broadhurst, ``Cosmic structure as the quantum interference of a coherent dark wave,''
Nature Phys. {\bf 10}, 496-499 (2014) [arXiv:1406.6586 [astro-ph.GA]]

\bibitem{Schive:2014hza} 
H.Y.~Schive, M.H.~Liao, T.P.~Woo, S.K.~Wong, T.~Chiueh, T.~Broadhurst and W.Y.P.~Hwang, ``Understanding the core-halo relation of quantum wave dark matter, $\psi$DM, from 3D simulations,'' 
Phys. Rev. Lett. {\bf 113}, 261302 (2014) [arXiv:1407.7762]

\bibitem{Schwabe:2016rze}
B.~Schwabe, J.C.~Niemeyer and J.F.~Engels, ``Simulations of solitonic core mergers in ultra-light axion dark matter cosmologies,'' 
Phys. Rev. D {\bf 94}, 043513 (2016) [arXiv:1606.05151]

\bibitem{Veltmaat:2016rxo}
J.~Veltmaat and J.C.~Niemeyer, ``Cosmological particle-in-cell simulations with ultra-light axion dark matter,''
Phys. Rev. D {\bf 94}, 123523 (2016) [arXiv:1608.00802]

\bibitem{Du:2016aik}
X.~Du, C.~Behrens, J.C.~Niemeyer and B.~Schwabe, ``The core-halo mass relation of ultra-light axion dark matter from merger history,'' 
Phys. Rev. D {\bf 95}, 043519 (2017) [arXiv:1609.09414]

\bibitem{Gonzalez-Morales:2016yaf}
A.X.~Gonzalez-Morales, D.J.E.~Marsh, J.~Pe\~narrubia and L.A.~Ure\~na-L\'opez, ``Unbiased constraints on ultralight axion mass from dwarf spheroidal galaxies,''
Mon. Not. Roy. Astron. Soc. {\bf 472}, 1346--1360 (2017) [arXiv:1609.05856 [astro-ph.CO]]

\bibitem{Peccei77} 
R.D.~Peccei and H.R.~Quinn, ``CP conservation in the presence of pseudoparticles,'' 
Phys. Rev. Lett. {\bf 38}, 1440 (1977)

\bibitem{Preskill83}
J.~Preskill, M.B.~Wise and F.~Wilczek, ``Cosmology of the invisible axion,'' 
Phys. Lett. B {\bf 120}, 127 (1983) 

\bibitem{Abbott83}
L.F.~Abbott and P.~Sikivie, ``A cosmological bound on the invisible axion,'' 
Phys. Lett. B {\bf 120}, 133 (1983) 

\bibitem{Dine83}
M.~Dine and W.~Fischler, ``The not so harmless axion,'' 
Phys. Lett. B {\bf 120}, 137 (1983) 

\bibitem{Hertzberg08}
M.P~Hertzberg, M.~Tegmark and F.~Wilczek, ``Axion cosmology and the energy scale of inflation,'' 
Phys. Rev. D {\bf 78}, 083507 (2008)

\bibitem{Sikivie08}
P.~Sikivie, ``Axion Cosmology,'' 
Lect. Notes Phys. {\bf 741}, 19-50 (2008) [astro-ph/0610440]

\bibitem{Hu:2000ke}
W.~Hu, R.~Barkana and A.~Gruzinov,``Cold and fuzzy dark matter,''
Phys. Rev. Lett. \textbf{85}, 1158-1161 (2000) 
[arXiv:astro-ph/0003365].

\bibitem{Matos:2000ss}
T.~Matos and L.~A.~Ure\~na-L\'opez,``A Further analysis of a cosmological model of quintessence and scalar dark matter,''
Phys. Rev. D \textbf{63}, 063506 (2001)
[arXiv:astro-ph/0006024]

\bibitem{Arvanitaki:2009fg}
A.~Arvanitaki, S.~Dimopoulos, S.~Dubovsky, N.~Kaloper and J.~March-Russell, ``String axiverse,''
Phys. Rev. D {\bf 81}, 123530 (2010) [arXiv:0905.4720 [hep-th]]


\bibitem{Suarez:2013iw}
A.~Su\'arez, V.H.~Robles and T.~Matos, ``A review on the scalar field/Bose-Einstein condensate dark matter model,''
Astrophys. Space Sci. Proc. {\bf 38}, 107--142 (2014) [arXiv:1302.0903 [astro-ph.CO]]

\bibitem{Marsh:2015xka} D.J.E.~Marsh, ``Axion cosmology,'' 
Phys. Rept. {\bf 643}, 1-79 (2016) [arXiv:1510.07633 [astro-ph.CO]]

\bibitem{Hui:2016ltb} L.~Hui, J.P.~Ostriker, S.~Tremaine, E.~Witten, ``Ultralight scalars as cosmological dark matter,''
Phys. Rev. D {\bf 95}, 043541 (2017) [arXiv:1610.08297 [astro-ph.CO]]

\bibitem{Niemeyer:2019aqm}
J.C.~Niemeyer, ``Small-scale structure of fuzzy and axion-like dark matter,'' 
Prog. Part. Nucl. Phys. {\bf 113}, 103787 (2020) [arXiv:1912.07064]

\bibitem{Luis2019}
 L.A.~Ure\~na-L\'opez, ``Brief review on scalar field dark matter models,''
 Front. Astron. Space Sci. {\bf 12}, (2019) [https://doi.org/10.3389/fspas.2019.00047]
 
\bibitem{Ferreira:2020fam} 
E.G.M.~Ferreira, ``Ultra-light dark matter,''
Astron. Astrophys. Rev. {\bf 29}, 7 (2021) [arXiv:2005.03254 [astro-ph.CO]]
 
\bibitem{Ho:2002vz} J.~Ho, F.~C.~Khanna, C.~H.~Lee, ``Boson stars with self-interacting quantum scalar fields,'' (2003) [arXiv:gr-qc/0207073] 

\bibitem{Matos2007} T.~Matos and L.A.~Ure\~na-L\'opez, ``Flat rotation curves in scalar field galaxy halos,''
Gen. Rel. Grav. {\bf 39}, 1279-1286 (2007) 

\bibitem{Bernal:2009zy} 
A.~Bernal, J.~Barranco, D.~Alic and C.~Palenzuela, ``Multi-state boson stars,''
Phys. Rev. D {\bf 81}, 044031 (2010) [arXiv:0908.2435 [gr-qc]] 

\bibitem{UrenaLopez:2010ur}
L.A.~Une\~na-L\'opez and A.~Bernal, ``Bosonic gas as a galactic dark matter halo,''
Phys. Rev. D {\bf 82}, 123535 (2010) [arXiv:1008.1231 [gr-qc]]

\bibitem{Barranco:2010ib}
J.~Barranco and A.~Bernal, ``Self-gravitating system made of axions,''
Phys. Rev. D {\bf 83}, 043525 (2011) [arXiv:1001.1769 [astro-ph.CO]]

\bibitem{Berczi:2020nqy}
B.~Berczi, P.M.~Saffin and S.Y.~Zhou, ``Gravitational collapse with quantum fields,''
Phys. Rev. D {\bf 104}, L041703 (2021) [arXiv:2010.10142 [gr-qc]]

\bibitem{Berczi:2021hdh}
B.~Berczi, P.M.~Saffin and S.Y.~Zhou, ``Gravitational collapse of quantum fields and Choptuik scaling,''
JHEP {\bf 02}, 183 (2022) [arXiv:2111.11400 [hep-th]]

\bibitem{Guenther:2020kro}
J.~Guenther, C.~Hoelbling and L.~Varnhorst, ``Semiclassical gravitational collapse of a radially symmetric massless scalar quantum field,''
Phys. Rev. D {\bf 105}, 105010 (2022) [arXiv:2010.13215 [gr-qc]]

\bibitem{arrechea2021semiclassical} J.~Arrechea, C.~Barcel\'o, R.~Carballo-Rubio and L.~J.~Garay, ``Semiclassical relativistic stars,'' 
Sci. Rep. {\bf 12}, 15958 (2021)
[arXiv:2110.15808 [gr-qc]] 

\bibitem{Verdaguer2020}
B.-L.~Hu and E.~Verdaguer, {\it Semiclassical and Stochastic Gravity,} 
Cambridge University Press (2020) 599pp 

\bibitem{semiclassical}
See e.g. Birrelll and Davies~\cite{Birrell1984} Chapter~6 ``Stress-tensor renormalization''; Wald~\cite{Wald1994} Section~4.6 ``The stress-energy tensor''; 
Mukhanov and Winitzki~\cite{Mukhanov2007} Part~2 ``Path integrals and vacuum polarization''; and Parker and 
Toms~\cite{Parker2009} Chapter~3 ``Expectation values quadratic in fields''.
See also Ref.~\cite{Burgess:2003jk} for a general discussion on quantum gravity at accessible scales.

\bibitem{Birrell1984}  
  N.~D.~Birrell and P.~C.~W. Davies, {\it Quantum Fields in Curved Space,} 
Cambridge University Press (1984) 352pp

\bibitem{Wald1994}
  R.~M.~Wald, {\it Quantum Field Theory in Curved Spacetime and Black Hole Thermodynamics,}
University Of Chicago Press (1994) 220pp

\bibitem{Mukhanov2007}
  V.~Mukhanov and S.~Winitzki, {\it Introduction to Quantum Effects in Gravity,}
Cambridge University Press (2007) 284pp 

\bibitem{Parker2009}
  L.~Parker and D.~Toms, {\it Quantum Field Theory in Curved Spacetime: Quantized Fields and Gravity,} 
Cambridge University Press (2009) 472pp 

\bibitem{Burgess:2003jk}
C.P.~Burgess, ``Quantum gravity in everyday life: General relativity as an effective field theory,''
Living Rev. Rel. {\bf 7}, 5--56 (2004) [arXiv:gr-qc/0311082]

\bibitem{Jordan:1986ug} 
R.D.~Jordan, ``Effective field equations for expectation values,''
Phys. Rev. D {\bf 33}, 444-454 (1986)

\bibitem{Calzetta:1986ey}
E.~Calzetta and B.L.~Hu, ``Closed time path functional formalism in curved space-time: Application to cosmological back reaction problems,''
Phys. Rev. D {\bf 35}, 495 (1987)

\bibitem{Paz:1990jg}
J.P.~Paz, ``Anisotropy dissipation in the early universe: Finite temperature effects reexamined,''
Phys. Rev. D {\bf 41}, 1054-1066 (1990)

\bibitem{Campos:1993ug}
A.~Campos and E.~Verdaguer, ``Semiclassical equations for weakly inhomogeneous cosmologies,''
Phys. Rev. D {\bf 49}, 1861-1880 (1994)

\bibitem{DiezTejedor2012}
A.~Diez-Tejedor and D.~Sudarsky,
``Towards a formal description of the collapse approach to the inflationary origin of the seeds of cosmic structure'', JCAP, {\bf 07}, 045 (2012)

\bibitem{Parker1974} 
L.~Parker and S.A.~Fulling, ``Adiabatic regularization of the energy momentum tensor of a quantized field in homogeneous spaces,'' 
Phys. Rev. D {\bf 9}, 341-354 (1974)

\bibitem{Pauli1940} 
W.~Pauli and F.~Villars, ``On the invariant regularization in relativistic quantum theory,'' 
Rev. Mod. Phys. {\bf 21}, 434-444 (1949) 

\bibitem{Weinberg:2010wq}
S.~Weinberg, ``Ultraviolet divergences in cosmological correlations,''
Phys. Rev. D {\bf 83}, 063508 (2011) [arXiv:1011.1630 [hep-th]]

\bibitem{Armendariz-Picon:2019csc} 
C.~Armendariz-Picon, ``On the expected production of gravitational waves during preheating,'' 
JCAP {\bf 08}, 012 (2019) [arXiv:1905.05724 [astro-ph.CO]].

\bibitem{Lieb77}
E.H. Lieb, ``Existence and Uniqueness of the Minimizing Solution of Choquard's Nonlinear Equation", Studies in Applied Mathematics {\bf 57} (1977) 93--105.

\bibitem{KavianMuschler2015}
O. Kavian and S. Mischler, ``A global approach to the Schr\"odinger-Poisson system: An existence result in the case of infinitely many states", J. Math. Pures Appl. {\bf 104} (2015) 942-964.

\bibitem{Alcubierre:2018ahf}  
M.~Alcubierre, J.~Barranco, A.~Bernal, J.C.~Degollado, A.~Diez-Tejedor, M.~Megevand, D.~N\'u\~nez and O.~Sarbach, ``$\ell$-Boson stars,''
Class. Quant. Grav. {\bf 35}, 19LT01 (2018) [arXiv:1805.11488 [gr-qc]]

\bibitem{Alcubierre:2019qnh}
M.~Alcubierre, J.~Barranco, A.~Bernal, J.C.~Degollado, A.~Diez-Tejedor, M.~Megevand, D.~N\'u\~nez and O.~Sarbach, ``Dynamical evolutions of $\ell$-boson stars in spherical symmetry,''
Class. Quant. Grav. {\bf 36}, 215013 (2019) [arXiv:1906.08959 [gr-qc]]

\bibitem{Alcubierre:2021mvs}
M.~Alcubierre, J.~Barranco, A.~Bernal, J.C.~Degollado, A.~Diez-Tejedor, M.~Megevand, D.~N\'u\~nez and O.~Sarbach, ``On the linear stability of $\ell$-boson stars with respect to radial perturbations,''
Class. Quant. Grav. {\bf 38}, 174001  (2021) [arXiv:2103.15012 [gr-qc]]

\bibitem{Alcubierre:2021psa}
  M.~Alcubierre, J.~Barranco,  A.~Bernal, J.~C.~Degollado, A.~Diez-Tejedor, V.~Jaramillo, M.~Megevand, D.~N\'u\~nez, O.~Sarbach, ``Extreme \ensuremath{\ell}-boson stars,''
Class. Quant. Grav. {\bf 39}, 094001 (2022) [arXiv:2112.04529 [gr-qc]]

\bibitem{Jaramillo:2022zwg}
  V.~Jaramillo, N.~Sanchis-Gual, J.~Barranco, A.~Bernal, J.~.C.~Degollado, C.~Herdeiro, M.~Megevand, D.~N\'u\~nez, ``Head-on collisions of \ensuremath{\ell}-boson stars,''
  Phys. Rev. D {\bf} 105, 104057 (2022) [arXiv:2202.00696 [gr-qc]] 
  
\bibitem{jaramillo19} V. Jaramillo P\'erez, {\it L\'imite de campo d\'ebil para el campo escalar autogravitante,} Master's Thesis, Universidad Nacional Aut\'onoma de M\'exico (2019). 

\bibitem{GU2020} F.~S.~Guzman and L.~A.~Ure\~na-L\'opez, ``Gravitational atoms: General framework for the construction of multistate axially symmetric solutions of the Schr\"{o}dinger-Poisson system,''
  Physical Review D {\bf 101}, 8 (2020) [arXiv:1912.10585 [astro-ph.GA]]

\bibitem{nambo21} E.~Ch\'avez Nambo, {\it Sobre la existencia de estrellas de bosones newtonianas con momento angular en simetr\'ia esf\'erica,}
Master's Thesis, Universidad Michoacana de San Nicol\'as de Hidalgo (2021)
  
\bibitem{Vicens:2018kdk}
J.~Vicens, J.~Salvado and J.~Miralda-Escude, ``Bosonic dark matter halos: excited states and relaxation in the potential of the ground state,''
(2018) [arXiv:1802.10513 [astro-ph.CO]]

\bibitem{Lin:2018whl}
S-C.~Lin, H-Y.~Schive, S-K.~Wong and T.~Chiueh, ``Self-consistent construction of virialized wave dark matter halos,''
Phys. Rev. D {\bf 97}, 103523 (2018) [arXiv:1801.02320 [astro-ph.CO]] 
 
\bibitem{Seidel91}
E.~Seidel and W.M.~Suen, ``Oscillating soliton stars,'' 
Phys. Rev. Lett. {\bf 66}, 1659 (1991)

\bibitem{Coleman:1985ki} 
S.R.~Coleman, ``Q-balls,'' 
Nucl. Phys. B {\bf 262}, 263 (1985) [Addendum: Nucl. Phys. B {\bf 269}, 744 (1986)]

\bibitem{DiGiovanni:2021vlu}
F.~Di Giovanni, S.~Fakhry, N.~Sanchis-Gual, J.~C.~Degollado and J.~A.~Font,
``A stabilization mechanism for excited fermion\textendash{}boson stars,''
Class. Quant. Grav. \textbf{38}, (2021) no.19, 194001
[arXiv:2105.00530 [gr-qc]]

\bibitem{Herdeiro:2022gzp}
C.~Herdeiro and E.~Radu, ``On the classicality of bosonic stars,'' 
Int. J. Mod. Phys. D {\bf 31}, 2242022 (2022) [arXiv:2205.05395 [gr-qc]]

\bibitem{Hartree1928}
D.R.~Hartree, ``The wave mechanics of an atom with a non-coulomb central field,'' 
Math. Proc. Camb. Philos. Soc. {\bf 24}, 111 (1928)

\bibitem{Guth:2014hsa}
A.H.~Guth, M.P.~Hertzberg and C.~Prescod-Weinstein, ``Do dark matter axions form a condensate with long-range correlation?,''
Phys. Rev. D {\bf 92}, 103513 (2015) [arXiv:1412.5930 [astro-ph.CO]]

\bibitem{Graham2013} 
P.W.~Graham and R.~Surjeet, ``New observables for direct detection of axion dark matter,''
Phys. Rev. D {\bf 88}, 035023 (2013) [arXiv:1306.6088 [hep-ph]]

\bibitem{Carney:2019cio}
D.~Carney, A.~Hook, Z.~Liu, J.M.~Taylor and Y.~Zhao, ``Ultralight dark matter detection with mechanical quantum sensors,''
New J. Phys. {\bf 23}, 023041 (2021) [arXiv:1908.04797 [hep-ph]]

\bibitem{Donohue:2021jbv}
C.M.~Donohue, S.~Gardner and W.~Korsch,  ``LC circuits for the direct detection of ultralight dark matter candidates,''
(2021) [arXiv:2109.08163 [hep-ph]]

\bibitem{Tsai:2021lly}
Y.-D.~Tsai, J.~Eby, and M.S.~Safronova, Marianna S, ``SpaceQ -- Direct detection of ultralight dark matter with space quantum sensors,''
(2021) [arXiv:2112.07674 [hep-ph]]

\bibitem{Perez:2005gh}
A.~Perez, H.~Sahlmann and D.~Sudarsky, ``On the quantum origin of the seeds of cosmic structure,''
Class. Quant. Grav. {\bf 23}, 2317--2354 (2006) [arXiv:gr-qc/0508100]

\bibitem{Sudarsky:2009za} 
D.~Sudarsky, ``Shortcomings in the understanding of why cosmological perturbations look classical,''
Int. J. Mod. Phys. D {\bf 20}, 509--552 (2011) [arXiv:gr-qc/0906.0315]

\bibitem{Aguirre:2015mva}
A.~Aguirre and A.~Diez-Tejedor, ``Asymmetric condensed dark matter,''
JCAP {\bf 04}, 019 (2016) [arXiv:1502.07354 [astro-ph.CO]]

\bibitem{Sikivie:2016enz}
P.~Sikivie and E.M.~Todarello, ``Duration of classicality in highly degenerate interacting Bosonic systems,''
Phys. Lett. B {\bf 770}, 331--334 (2017) [arXiv:1607.00949 [hep-ph]]

\bibitem{Hertzberg:2016tal}
M.P.~Hertzberg, ``Quantum and classical behavior in interacting bosonic systems,''
JCAP {\bf 11}, 037 (2016) [arXiv:1609.01342 [hep-ph]]

\bibitem{Allali:2020shm}
I.J.~Allali and M.P.~Hertzberg, ``Decoherence from General Relativity,''
Phys. Rev. D {\bf 103}, 104053 (2021) [arXiv:2012.12903 [gr-qc]]


\bibitem{Flanagan:1996gw}
E.E.~Flanagan and R.M.~Wald, ``Does back reaction enforce the averaged null energy condition in semiclassical gravity?''
Phys. Rev. D {\bf 54}, 6233-6283 (1996) [gr-qc/9602052]

\bibitem{Jacobson:2003vx}
T.~Jacobson, ``Introduction to quantum fields in curved space-time and the Hawking effect,''
contribution to {\it School on Quantum Gravity}, pages 39--89 (2003) [gr-qc/0308048 [gr-qc]]

\bibitem{Ashtekar:1975zn}
A. Ashtekar and A. Magnon, ``Quantum fields in curved space-times,''
Proc. Roy. Soc. Lond. A, {\bf 346}, 375--394, 1975.

\bibitem{Sakurai94}
 J.J.~Sakurai, {\it Modern Quantum Mechanics,}
Addison-Wesley Publishing Company (1994) 500pp 

\bibitem{Zhang1990}
W.M.~Zhang, D.H.~Feng and R.~Gilmore, ``Coherent states: Theory and some applications,''
Rev. Mod. Phys. {\bf 62}, 867 (1990)

\bibitem{Sanders2012}
B.C.~Sanders, ``Review of entangled coherent states,''
J. Phys. A: Math. Theor. {\bf 45}, 244002 (2012)


\bibitem{Glauber63}
R.J.~Glauber, ``Coherent and incoherent states of radiation field,'' 
Phys. Rev. {\bf 131}, 2766-2788 (1963)

\bibitem{Armendariz-Picon:2020tkc} C.~Armendariz-Picon, ``On the expected backreaction during preheating,'' JCAP {\bf 05}, 035 (2020) [arXiv:2003.01542 [gr-qc]]

\bibitem{Juarez-Aubry:2019jon} 
B.~A.~Juarez-Aubry, T.~Miramontes and D.~Sudarsky, ``Semiclassical theories as initial value problems,''
J. Math. Phys. {\bf 61}, 032301 (2020) [arXiv:1907.09960 [math-ph]]

\bibitem{bjuarez2021a} B.~A.~Juarez-Aubry, ``Semiclassical gravity in static spacetimes as a constrained initial value problem,'' 
Annales Henri Poincar\'e {\bf 23}, 4, 1451-1487 (2022) [arXiv:2011.05947 [gr-qc]]

\bibitem{bjuarez2021b} B.~A.~Juarez-Aubry and S.~L.~Modak, ``Semiclassical gravity with a conformally covariant field in globally hyperbolic spacetimes,'' 
J. Math. Phys. {\bf 63}, 9, 092303 (2022)  [arXiv:2110.01719 [math-ph]]

\bibitem{Juarez-Aubry:2022qdp}
B.~A.~Juarez-Aubry, B.~Kay, T.~Miramontes and D.~Sudarsky, ``On the initial value problem for semiclassical gravity without and with quantum state collapses,''
(2022) [arXiv:2205.11671 [gr-qc]]

\bibitem{Canate:2018wtx}
P.~Canate, E.~Ramirez and D.~Sudarsky, ``Semiclassical self consistent treatment of the emergence of seeds of cosmic structure. The second order construction,''
JCAP {\bf 08}, 043 (2018); Erratum: JCAP 10, E01 (2018) [arXiv:1802.02238 [gr-qc]]


\bibitem{ReedSimonVol2}
M.~Reed and B.~Simon, {\it Methods of Modern Mathematical Physics, Vol. II: Fourier Analysis, Self-Adjointness}, Academic Press, San Diego (1980)

\bibitem{ReedSimonVol1}
M.~Reed and B.~Simon, {\it Methods of Modern Mathematical Physics, Vol. I: Functional Analysis}, Academic Press, San Diego (1980)

\bibitem{ReedSimonVol4}
M.~Reed and B.~Simon, {\it Methods of Modern Mathematical Physics, Vol. IV: Analysis of Operators}, Academic Press, San Diego (1980)  

\bibitem{Much2021}
A.~Much and R.~Oeckl, ``Self-Adjointness in Klein-Gordon theory on globally hyperbolic spacetimes,''
Math. Phys. Anal. Geom. {\bf 24}, 5 (2021) [arXiv:1804.07782 [math-ph]]

\bibitem{Alcubierre2008}
M.~Alcubierre, {\it Introduction to 3+1 Numerical Relativity,}
Oxford University Press (2008) 444pp

\bibitem{Mielke:1980sa}
E.W.~Mielke and R.~Scherzer, ``Geon type solutions of the nonlinear Heisenberg-Klein-Gordon equation,''
Phys. Rev. D {\bf 24}, 2111 (1981)

\bibitem{Andreasson-LivRev}
H. Andr\'easson, ``The Einstein-Vlasov system/kinetic theory,''
Living Rev. Rel. {\bf 14}, 4 (2011) [arXiv:1106.1367 [gr-qc]]
  
\bibitem{MorozPenroseTod1998}
I.M. Moroz, R. Penrose and P. Tod, ``Spherically-symmetric solutions of the {S}chr\"odinger-{N}ewton equations", Class. and Quantum Grav. {\bf 15}, (1998), 2733--2742.
   
\bibitem{DeMartino:2018zkx}
I.~De~Martino, T.~Broadhurst, S.-H.H.~Tye, T.~Chiueh, and H.-Y.~Schive, ``Dynamical evidence of a solitonic core of $10^{9}M_\odot$ in the Milky Way,''
Phys. Dark Univ. {\bf 28}, 100503 (2020) [arXiv:1807.08153 [astro-ph.GA]]

\bibitem{Atlas3D} Pierre-Alan Duc {\it et. al}. ``The Atlas 3D project XXIX. The new look of early-type galaxies and surrounding fields disclosed by extremely deep optical images,''
MNRAS {\bf 446}, 120 (2015) [arXiv:1410.0981 [astro-ph.GA]]
  
  
  
\end{thebibliography}
\end{document}